\documentstyle[12pt,epsfig]{article}
\newcommand{\be}{\begin{equation}}
\newcommand{\ee}{\end{equation}}
\newcommand{\bea}{\begin{eqnarray}}
\newcommand{\eea}{\end{eqnarray}}
\newcommand{\beas}{\begin{eqnarray*}}
\newcommand{\eeas}{\end{eqnarray*}}
\newcommand{\bi}{\begin{itemize}}
\newcommand{\ei}{\end{itemize}}
\newcommand{\bc}{\begin{center}}
\newcommand{\ec}{\end{center}}
\newcommand{\bfl}{\begin{flushleft}}
\newcommand{\efl}{\end{flushleft}}
\newcommand{\bfr}{\begin{flushright}}
\newcommand{\efr}{\end{flushright}}
\newcommand{\f}{\frac}

\def\6{\partial} \def\a{\alpha} \def\b{\beta}
\def\g{\gamma} \def\d{\delta} 
\def\e{\epsilon}
\def\z{\zeta} \def\h{\eta} 
 \def\k{\kappa} \def\l{\lambda}
\def\m{\mu} \def\n{\nu}  \def\p{\pi}
\def\r{\rho} \def\s{\sigma} \def\t{\tau}
  
 \def\G{\Gamma} 
 \def\L{\Lambda}

\def\ti{\tilde} 
\def\non{\nonumber\\}
\def\fr{\frac}
\def\le{\left}
\def\rh{\right}
\def\fr{\frac}

\begin{document}

\title{Introductory Lectures on D-Branes}

\author{Ion Vasile Vancea\footnote{On leave from
Babes-Bolyai University of Cluj}\\
Instituto de F\'{\i}sica Te\'{o}rica, Universidade Estadual Paulista\\
Rua Pamplona 145, 01405-900, S\~{a}o Paulo SP, Brasil\\
Email: ivancea@ift.unesp.br}

\maketitle

\abstract{This is a pedagogical introduction to $D$-branes, addressed  
to graduate students in field theory and particle physics and to other 
beginners in string theory. I am
not going to review the most recent results since there are already
many good papers on web devoted to that. Instead, I will present some
old techniques in some detail in order to show how
some basic properties of strings and branes as 
the massless  spectrum of string, the effective action of $D$-branes 
and 
their tension can be computed using QFT techniques. Also, I will 
present shortly 
the boundary state description of $D$-branes. The details are exposed 
for bosonic branes since I do not assume any previous knowledge of
supersymmetry which is not a requirement for this school. However, for 
completeness and to provide basic notions for other lectures, 
I will discuss the some properties of supersymmetric branes. 
The present lectures were delivered at Jorge Andr\'{e} Swieca School on 
Particle and 
Fields, 2001, Campos do Jord\~{a}o, Brazil.}

\newpage

\section{Introduction}

The fundamental problem of high energy theoretical physics is to 
provide us with a description of the intimate structure and interaction of the Nature. 
At low energies, accessible to particle physics experiments, 
the accepted models are based on QFT which predicts particle-like excitations 
of the fields interacting through three fundamental forces: electro-magnetic,
weak and strong and through a classical gravitational interaction. Nowadays, it
is known that the first three interactions can be unified at intermediate 
energies and it is believed that at a scale of energy around $10^{-33}$ cm, 
(the {\em Planck length}) all the interactions should be consistently 
described by an unique theory which is unknown yet. There are several arguments
in support of the idea that at this scale the excitations (or at least part of
them) of the fundamental object of the theory should be string-like rather than
point-like and studying such of hypothesis is the object of the string theory. 
N. Berkovits will review in his lectures at this school the problems of field 
theory and gravity at the Planck scale and the arguments in favor of string
theories. For our purpose, it is enough to say that many people consider
string (based) theories as the most promising candidate for the final theory
of the Nature basically because they predict all particles
and interactions among them, including the gravity, from a basic object 
which is the string. However, due to the largeness of the unification scale,
the predictions and the constructions are theoretical; all what is required is
that the theory satisfy some internal consistency conditions and reproduce at
the low energy limit the known theories, i.e. QFT and gravity. Results that 
could be 
decisively connected 
with phenomenology have not been rigorously obtained yet.  

Despite many of its conceptual successes, string theories suffer from 
several serious problems which serve as arguments in favor of criticism 
against them. One of the drawbacks of strings is that they appear 
in a too large 
number of theories, all with equal rights to a theory of 
everything: Type I, Type IIA, Type IIB, 
Heterotic $SO(32)$ and Heterotic $E_8 \times E_8$. All of these theories 
are supersymmetric and their names are 
related to the number and type of supersymmetry and of the gauge group. 
Another problem is that all these theories
live in a ten dimensional space-time with one time-like direction. 
Since the world in which we live has only three
space-like directions, one has to explain how the space-time of string 
theory reduces to the physical one. One way
of thinking to the dimensional reduction is by considering that some of the 
$d=10$ space-time directions are {\em compactified}.
There are some recent progress in this direction, but no completely 
satisfactory answer is known at present. However, with the discovery of the
D-branes six years ago, many revolutionary ideas about string theories and
new angles of attack to the old problems have been emerging.

The $D$-branes are extended physical objects discovered in string theories. 
One type of $D$-branes, called BPS-branes,
which saturate a relation between supersymmetry and energy, play an important 
role in establishing relations
among the string theories called {\em dualities}. By a duality, two string 
theories are mapped into one another. Since
the theories can describe the interactions in different regimes 
(for example on different compact space-time or
in weak-strong coupling limits), they can represent relations among 
these different limits (of string theories.)
This points out towards an unique underlying theory of which limits are 
described by string theories. This unknown yet
theory is called $M$ (or, sometimes $U$)-theory and one of its limits 
is the supergravity in $d=11$.  
Beside their importance in unification of string theories, the $D$-branes are 
necessary for their internal consistency. Indeed,
in Type II theories there are some massless p-form bosonic fields called 
Ramond-Ramond (RR) fields after the name of 
the perturbative sector of string spectra in which they appear. 
The $D$-brane are the sources of these fields and carry 
their charges.In simple situations in which the branes can be considered as 
hyperplanes, there are similar relations
between RR-charges and the topological and the Noether charges encountered in 
electro-magnetism. However, being extended objects, the $D$-branes can have 
more sophisticated topologies, in which case there are topological 
contributions to the charges. Therefore, the classification of charges is 
given by an extended cohomology called {\em K-theory} rather than by the 
De Rham cohomology. There are subtle relations in K-theory among 
different $D$-branes and between branes and $d=11$ supergravity. BPS as well 
as non-BPS branes are involved in that and the tachyon fields  
represent the mechanism that controls the evolution of the system.

Beside their crucial role in understanding string theories, the $D$-branes 
have been successfully used to explain 
various supersymmetric and non-supersymmetric field theories, the entropy of 
some models of black-holes and,
more recently, there have been attempts in understanding some cosmological 
issues and the hierarchy problem. New
developments appear daily on the net, and one of the hot topics is the 
geometrical structure of space-time and
field theories at the Planck scale, which could be non-commutative.

The $D$-branes represent a quickly developing field, where many novelties 
appear almost each year. There are many
fascinating problems in this direction which has been inspiring in both 
physics and mathematics. However, the purpose
of these lecture notes is by far more modest. I am not going to review nor 
the successes neither the various theories
that are based on and that involve the $D$-branes. To this end, there are many
very good references at http://xxx.lanl.gov. 

The aim of these lecture is to show to particle physics theorists how some
basic properties of D-branes have been computed and how field theory 
techniques can be used to obtain information about branes. By that, I hope 
that graduate students and researchers will feel more confident and 
encouraged to read string literature. Also, I would like to provide a 
background for the more advanced topics in brane theory that will be 
presented at this school by I. Ya. Arafeva, S. Minwalla and K. Stelle.  

Due to the pedagogical line that I had chosen and since most of the audience 
was not familiar with string theory, supersymmetry and supergravity, I 
focused in my talks on {\it physical properties} of bosonic $D$-branes as  
as low energy effective action, tension and boundary field description. 
However, some properties of supersymmetric branes also emerged mostly during
the discussion sessions, therefore I added a part in which the same
basic properties but for the supersymmetric branes are listed.

The structure of these notes is as follows. In Section 2 the basic results of 
free bosonic string theory are 
revisited. I will present the massless spectrum of either open and closed 
bosonic string theory to argue the
presence of graviton and other bosonic excitations which will be used in the 
next sections. In Section 3 we
deduce the Born-Infeld action or the low energy action of a $D$-brane. 
I work out in detail the case of the 
$D25$-brane and the sigma model in a general closed string background. 
This section is based on 
In Section 4 we find the tension of the $D$-branes. In Section 5 we present 
the microscopic description of
$D$-branes known as the boundary state formalism. In the last Section the
same results for supersymmetric branes are presented.

There are many references on strings and D-branes and
it would be impossible to mention all of them. I have tried to refeer 
mainly to more advanced lecture notes rather than original papers for 
pedagogical reasons. The references that I have been 
used throughout this work have been selecting according to my preference.

I would like to thank to M. C. B. Abdalla, N. Berkovits, C. T. Echevarria, 
A. L. Gadelha, V. D. Pershin, V. O. Rivelles, W. P. de Souza,
K. Stelle, B. Vallilo and to all those that have been attended these lectures 
for their stimulating discussions. Also, I am grateful to H. Kogetsu who pointed 
out an error in the first version of the paper. 

\newpage

\section{Review of Basic Results from Free Bosonic String Theory}

In this section I am going to review some basic results from free bosonic
string theory that will be useful in studying the D-branes. This is a
text-book material \cite{gsw,jp,cj} but it is included here since 
these ideas might be unfamiliar to some part of the audience.

\subsection{Classical free bosonic string theory}

The starting point in discussing a classical free string is its action which is
proportional to the area described by the string during its evolution
in spacetime \cite{gsw,jp,nb}. The classical action can be cast in a 
polynomial form known as Polyakov action which, 
in the conformal gauge and in the Minkowski background, is given by the 
following formula 
\be
S = - \frac{T_s}{2} \int_{\Sigma} d^2 \sigma \6_{\a}X^{\m}\6^{\a}X_{\m}.
\label{polyakov}
\ee
Here, $T_s$ is the string tension which is related to the Regge slope by the
formula $T_s = (2 \pi \a ')$ and  $X^{\m}$ are the string coordinates in
$D=26$ dimensional space-time, $\m =0, 1, \ldots, 25$. The world-sheet is
parametrized by a time-like parameter $\tau = \sigma^0 $ and a space-like one
$\sigma = \sigma^1 \in [0,\pi]$. The two metrics, on the world-sheet $\Sigma$
and on the space-time, respectively, are taken to be Minkowskian. 
The action given by
Eq.(\ref{polyakov}) has the following symmetries: $SO(1,25)$ in space-time, 
$SO(1,1)$ on the world-sheet and a residual two-dimensional
conformal symmetry \cite{gsw,jp,nb}.

The variational principle applied to the action (\ref{polyakov}) gives
the following equations of motion
\be
\6^{\a}\6_{\a}X^{\m}(\tau \s)=0
\label{eqmotion}
\ee
which hold only if the boundary conditions are satisfied, too. The closed
string boundary conditions simply express the fact that the string
coordinates are univalued
\be
X^{\m}(\tau,\sigma + \pi) = X^{\m}(\tau,\pi).
\label{boundclosed}
\ee
In the case of the open string, one can impose Poincar\'e invariant or
Poincar\'e
breaking boundary conditions. They are given by the following relations
\bea
\6_nX^{\m}|_{\6\Sigma} &=& 0 \\
\delta X^{\m}|_{\6\Sigma} &=& 0,
\label{boundopen}
\eea
which are known as Neumann b.c. and Dirichlet b.c., respectively. 
The topology of the world-sheet depends on the topology of the string. 
At classical level, the world-sheet of the closed string is equivalent 
to a cylinder or a two-dimensional annulus, while the  world-sheet of an 
{\em open} string can be continuously 
transformed into a disk with the boundary corresponding to the circle.

In order to determine the solutions of the equations of motion one can
study the variation of the action with fixed string configurations at the initial
and final time: $\delta X^{\m}(\tau_1) = \delta X^{\m}(\tau_2) = 0$. (The
boundary conditions may affect the finiteness of the physical quantities
derived out of the action but not the solutions of the equations of motion.) For these
configurations the boundary conditions are given by the following relations
\bea
\6_{\sigma} X^{\m}|_{\sigma =0,\pi} & = & 0~~~~~~~~ \mbox{N.b.c.}\\
X^{\m}|_{\sigma =0,\pi} & = & \mbox{ct.} ~~~~~~\mbox{D.b.c.}
\label{boundvar}
\eea

There is a constraint in the theory, namely that the energy-momentum tensor 
vanishes. This constraint can be understood by considering the string 
coupled to a two-dimensional graviton, i.e. on a curved world-sheet, 
where it arises as
the equation of motion of the two-dimensional gravitational potential.
However, in the present case it is a condition that should be imposed on
the system by hand \cite{gsw,jp,nb} and it is give by the relation
\be
T_{\a\b}=\6_\a X^\mu \6_\b X_\mu - 
\frac {1}{2}  \eta_{\a\b} \6_\g X^\mu \6^\g X_\mu = 0.
\label{energmom}  
\ee
The Weyl invariance of the action $h_{\a\b} \rightarrow \exp{\Lambda}~h_{\a\b}$
implies that the trace of the energy-momentum tensor vanishes 
\be
\mbox{Tr}~T_{\a\b} = T^{\a}_{\a}=0.
\label{vantrace}
\ee
The relations given by Eq.(\ref{energmom}) and Eq.(\ref{vantrace}) represent
strong constraints on the system and in the quantization process they should be
implemented at the quantum level. They reflect the fact that string theory is
a conformal field theory in two dimensions, a property that determines  all
the features of bosonic string physics.

The solutions of the equations of motion  can be
found by employing the method of separation of variables. The Fourier
expansion of closed string solution is given by 
\be
X^\mu(\t,\s)= x^{\mu} 
+ 2\a ' p^\mu \t +
i\sqrt{\f{\a '}{2}}
\sum _{n\neq 0}\frac{1}{n}(\a^{\mu}_{n}e^{-2in(\t - \s)}
+ {\tilde{\a}}^{\mu}_{n}e^{-2in(\t + \s)}).          
\label{closesol}
\ee
The above relation shows that the most general closed string solution is
a linear superposition of right- and left-moving modes with Fourier
coefficients $\a^{\mu}_{n}$ and ${\tilde{\a}}^{\mu}_{n}$, respectively.
Here $ x^{\mu}$ and $p^\mu $ represent the coordinates  of the center of 
mass of the closed string and their canonical conjugate momenta, respectively.

The boundary conditions of the open string can be chosen either Dirichlet or
Neumann on each direction. Therefore, one can have NN, DD, ND and DN solutions
of the equations of motion given by
\bea
\mbox{N-N}~&:&~~X^\mu(\t,\s)=x^{\mu} +2\a ' p^\mu \t +
2i\a '\sum _{n\neq 0}\frac{1}{n}\a^{\mu}_{n}e^{-in\t}\cos{n\s},          
\label{NNsol}\\
\mbox{D-D}~&:&~~X^\mu(\t,\s)=\f{x^{\mu}(\pi-\sigma)+ y^{\m}\sigma}{\pi} 
-
i\sqrt{2\a '}p^\mu\sum _{n\neq 0}\frac{1}{n}\a^{\mu}_{n}e^{-in\t}\sin{n\s}, 
\label{DDsol}\\
\mbox{D-N}~&:&~~
X^\mu(\t,\s)=x^{\mu} - 
i\sqrt{2\a '}\sum_{r \in \cal{Z}'}
\frac{1}{r}\a^{\mu}_{r}e^{-ir\t}\sin{r\s},          
\label{DNsol}\\
\mbox{N-D}~&:&~~ 
X^\mu(\t,\s)=y^{\mu} +
i\sqrt{2\a '}\sum_{r \in \cal{Z}'}
\frac{1}{r}\a^{\mu}_{r}e^{-ir\t}\cos{r\s},          
\label{NDsol}
\eea
where $\cal{Z}' = \cal{Z}$$+1/2$.
The closed string solution (\ref{closesol}) is Lorentz invariant in $D=26$ 
dimensions. The only open string solution which is invariant 
under
$SO(1,25)$ is (\ref{NNsol}). The other solutions break the Lorentz invariance 
down to $SO(1,p) \times SO(25-p)$.

\subsection{Massless spectrum of bosonic open string}

Let us see what is the particle content of the free bosonic string theory. To
this end one has to quantize the string, but care should be taken 
since the theory is subjected to the constraints (\ref{energmom}) and 
(\ref{vantrace}). Consequently,  not all degrees of freedom are physical. 
By implementing the constraints at the 
quantum level one can remove the effect of the non-physical degrees of 
freedom. One way of doing that is by employing the canonical quantization 
method which implies that the  constraints will appear 
as operatorial equations 
in the Fock space of the theory. Their solutions represent the physical 
states (for more details on the quantization of the string theory through
various methods see \cite{gsw}.) 

Consider the Lorentz invariant solution of the open string theory
((\ref{DDsol}), (\ref{DNsol}) and (\ref{NDsol}) are quantized in exactly the 
same way.) The coordinates $X^{\m}$ can be viewed as two-dimensional fields,
which have the equal-time commutators given by
\be
[X^{\m}(\t,\s),{\dot{X}}^\n (\t,\s ')] = i \eta^{\m\n}\delta(\s -\s ')
\label{eqtcomm1}
\ee
\be
[X^{\m}(\t,\s),X^\n (\t,\s ')] =
[{\dot{X}}^{\m}(\t,\s),{\dot{X}}^\m (\t,\s ')] = 0.
\label{eqtcomm2}
\ee
The $\eta^{00}$ component of the Minkowski metric generates negative norm,
unphysical states which must be removed from the spectrum. One can obtain
the commutation relations among the Fourier coefficients by plugging 
(\ref{NNsol}) into (\ref{eqtcomm1}) and (\ref{eqtcomm2}). The result is
\be
[\a^{\m}_{m},\a^{\n}_n] = m\eta^{\m\n}\delta_{m+n,0}
\label{alphacomm1}
\ee
\be
[x^\m,p^\n] = i\eta^{\m\n}.
\label{alphacomm2}
\ee
Note that for $\m \neq 0$ and $\n \neq 0$ the relations above are the usual
commutation relations of linear oscillators scaled by factor of $m$ from 
which we see that 
$\a_n$ operators with $ n > 0$ play the role of annihilation operators while
for $n < 0$ they act as creation operators. Therefore, it is possible to 
define 
a vacuum state with respect to these operators and to construct the Fock space.
There is no $n=0$ mode but one
define it as being the momenta of the center of mass \cite{gsw,jp,nb}.

The unphysical states corresponding to the time-like direction of space-time
are removed by the energy-momentum constraints. In order to implement them on
the Fock space, one has to write firstly the Fourier expansion of the 
components of the energy-momentum tensor \cite{gsw}. The Fourier coefficients
of $T_{\a\b}$ can be expressed in terms of Fourier coefficients of 
coordinates as follows
\bea
L_m & = & \f{1}{2}\sum_{n\neq 0} \a_{-n} \cdot \a_{n+m}\nonumber\\
L_0 & = & \a 'p^2 + \sum_{n=1}^{\infty}\a_{-n}\cdot\a_{n},
\label{viroperat}
\eea
where ``$\cdot$'' denotes the scalar product in the Minkowski space-time.
The Fourier coefficients of the energy-momentum tensor $L_m$ satisfy an 
algebra called the {\it Virasoro algebra} which has the following form
\be
[L_m,L_n ]= i(m-n)L_{m+n}.
\label{viralgebra}
\ee
The Virasoro algebra is the infinite algebra of the generators of the 
two-dimensional conformal group. The existence of this symmetry  guarantees 
that the system is integrable. However, upon quantization one can see that
an anomalous term appears in the algebra (\ref{viralgebra}). This term 
cancels if $D=26$.  

The physical states form a subspace of the Fock space which is defined by 
acting on the
vacuum with the creation operators. The vacuum $|0>$ is defined by the 
following conditions
\bea
|0> & \equiv & |0>_\a |p>\\
\a^{\m}_{n}|0>_\a & = & 0 ~,~~~~~~~~~~n>0 \\
{\hat{p}}^\m |p> & = & p^\m |p>.
\label{vacopen}
\eea
Alternatively, we will use the notation $|k>$ for $|0>$. The first two 
relations just define the vacuum of the linear oscillators. The last relation
tells that the vacuum continues to make sense when it is translated.

The physical states are defined as being those states of the Fock space that
obey the constraints coming from the vanishing energy-momentum tensor. One can
show that only half of the Virasoro operators should be imposed on the Fock
space since the full set of constraints is incompatible with the Virasoro 
algebra \cite{gsw,jp,cj}. The conditions that define the physical states are
\bea
L_m |\phi> & = & 0~,~~~~~m>0\\
(L_0 -1)|\phi> & = & 0.
\label{physstates}
\eea
The $-1$ term in the last equation above comes from the normal ordering of the
operator $L_0$. Since the Fourier coefficients of the energy-momentum tensor 
$L_m$ are expressed in terms of creation and annihilation operators as 
Eq.(\ref{viroperat}) shows, one has to normal order them through the 
quantization process. The normal ordering affects only $L_0$ operator by the
$-1$ term as can be easily checked up. 

A special class of states are the {\em spurious states}. These are states that
belong to the equivalence classes of physical states. Two physical states are
said to be equivalent if
\be
|\phi '> \sim |\phi> ~\Leftrightarrow ~ \exists 
~|\psi>~~:~~|\phi '> = |\phi> + |\psi>
\label{equivstates}
\ee
and $|\psi>$ is a spurious state, i.e. it satisfies the following equations:
\bea
(L_0 -1)|\psi> &=& 0 \\
<\psi|\phi> &=&0, 
\label{spurstate}
\eea
for any physical state $|\phi>$. An important example of a spurious state is the
one obtained by the action of the operator $L_{-1}$ on the vacuum: 
\be
|\psi> = L_{-1} |0>,
\label{spurphot}
\ee
 which is used to show the gauge invariance of the vector 
state as we will see later. Any state that is a linear combination 
of $L_{-n}$ operators is a spurious state
\be
|\psi>= \sum_m^{\infty}a_m L_{-m} |\chi_m>.
\label{spurstate1}
\ee 

The physical states can be classified according to their mass. The masses are 
the eigenvalues of the mass operator related to the momentum of the particle 
by the relativistic mass-shell relation: $M^2 = -p^2$. The mass operator is 
given by the action of $L_0 -1$ which describes the mass-shell operator. 
Indeed, from 
\be
(\a' p^2 + \sum_{n=1}^{\infty} \a_{-n}\cdot \a_n -1)|\phi>=0
\label{deducemass}
\ee
it follows that the mass operator is given by the formula
\be
M^2 = \f{1}{\a '}(\sum_{n=1}^{\infty}\a_{-n}\cdot\a_n -1).
\label{massoperat}
\ee 

Let us compute the masses of the first two levels. In the vacuum state there
are no contributions from the oscillators and one can see from 
Eq.(\ref{massoperat}) that the mass of the vacuum is negative
\be
m^2 = - \f{1}{\a '},
\label{vacuummass}
\ee
where $m^2$ denotes the eigenvalue of the mass operator. Therefore, the vacuum
is a tachyon. The tachyon travels faster than light and can be excited to any
negative energy, thus making the theory unconsistent. The presence of tachyon 
shows that the chosen background of the string theory (i.e. bosonic string), if
existed, is unstable.

The next states are obtained obtained from vacuum by acting with the
creation operators on $|0>$
\be
\a^{\m}_{-1}|0>
\label{photstates}.
\ee 
Due to the commutation relations given by Eq.(\ref{alphacomm1}), the 
contribution of the oscillators is exactly $+1$ so that the mass of the
states (\ref{photstates}) is $m^2 = 0$. They are states in the massless
representation of the group $SO(1,25)$. Therefore, their physical degrees of 
freedom are in the vector representation of SO(24) which is the little group
of the Lorentz group in $D=26$. If we construct the vector
\be
|\xi>=\xi_\m\a^{\m}_{-1}|0>,
\label{photon}
\ee  
we can see that there are equivalent vectors to it, namely
\be
|\xi '> = |\xi> + \lambda|\psi>,
\label{gaugetransf1}
\ee
where $|\psi>$ is the spurious state given in Eq.(\ref{spurphot}) and 
$\lambda$ is an arbitrary complex number. Thus, we can interpret 
Eq.(\ref{gaugetransf1}) as a $U(1)$ gauge transformation
\be
\xi^\m \rightarrow \xi^\m + \lambda k^\m,
\label{gaugetransf2}
\ee 
where $k^\m$ is the momentum of the spurious state. In conclusion, {\em 
the state $|\xi>$ describes a massless photon in D=26 with 24 transverse 
polarizations.}

\subsection{Massless spectrum of bosonic closed string}

The classical string has two types of oscillation modes: left and right. The
waves that propagate to the left are independent of the ones that propagate
to the right. Therefore, in the quantum theory the left and right Fock spaces
will be independent and the total Fock space of the closed string will be the
their tensor product. Also, the operators split into operators that act on
the left-modes and the ones that act on the right-modes, respectively.

Upon quantization of the solution given in Eq.(\ref{closesol}) the Fourier 
modes $\alpha$ and $\tilde{\a}$ become operators acting on the right and left
Fock spaces. Each of these two sets of operators obeys the algebra 
(\ref{alphacomm1}) and the two algebras are independent \cite{gsw,jp,nb}.
In order to define the physical states one has to impose the vanishing of the
energy-momentum tensor on the total Fock space. Moreover, since the string is
closed, there is and extra symmetry that should be maintained at the quantum 
level, namely the world-sheet invariance under the translation along $\s$.

The vacuum of the theory is defined as in the case of the 
open string
\bea
|0> &\equiv & |0>_\a |0>_{\tilde{\a}}|p>\\
\a^{\m}_{n} |0>_\a & = & 0 \\
{\tilde{\a}}^{\m}_{n} |0>_{\tilde{\a}} & = & 0~,~~~~~~~~~~~n > 0 \\
{\hat{p}}^\m |p> & = & p^\m |p>.
\label{vacclosed}
\eea
The physical states are defined by imposing half of the Virasoro operators on
the Fock space as well as the invariance of the world-sheet under the 
translation along $\s$ which is generated by the operator $L_0 - {\tilde{L}}_0$
\cite{gsw,jp}
\bea
L_m |\phi> &=& {\tilde{L}}_m |\phi> = 0 ~,~~~~~~~~~~~ m>0\\ 
(L_0 - 1)|\phi> &=& ({\tilde{L}}_0 - 1)|\phi> = 0\\
(L_0 - {\tilde{L}}_0) |\phi>& = &0. 
\label{physclosed}
\eea
The last condition above implies that the right- and left-modes should 
come in pairs of equal mass since the operator $L_0$ is related to $M^2$.

The spurious states are defined through the following relations
\bea
(L_0 -1) |\psi>&=&0\\
({\tilde{L}}_0 - 1) |\psi> &=& 0\\
<\psi|\phi> &=& 0\\ 
(L_0 - {\tilde{L}}_0)|\psi> &=&0,
\label{spurclosed}
\eea
where $|\psi>$ is an arbitrary physical state. The last relation guarantees
that the state to which a spurious state is added to remains invariant under
translation by $\s$.

One can find the mass operator as in the open string case or by observing
that the total mass should be the sum of the left and right mass operators
\bea
M^2 &=& M_{L}^2 + M_{R}^2\\
M^2 &=&\f{2}{\a '}(\sum_{n=1}^{\infty}(\a_{-n}\cdot \a_n + {\tilde{\a}}_{-n}
\cdot {\tilde{\a}}_n) -2).
\label{massclosed}
\eea

Let us look at the first states in the spectrum. The vacuum state is a tachyon
of mass
\be
m^2 = -\f{4}{\a '}
\label{masstachclosed}
\ee
with all the consequences that we saw in the case of the open string. The next 
states are constructed by applying equal number of creation operators on 
vacuum from left and right sectors as dictated by the level matching condition.
The first states are given by
\be
\a^{\m}_{-1}{\tilde{\a}}^{\n}_{-1}|0>
\label{masslesclosed}
\ee 
and it is easy to see that they are massless. In $D=26$ there are $26 \times 26$
such of massless states. They form a tensor in a reducible representation of
the group $SO(1,25)$. It splits into irreducible representation as follows
\[ 
\a^{\m}_{-1}{\tilde{\a}}^{\n}_{-1}|0> = 
\left\{ 
        \begin{array}{lll}
          \a^{(\m}_{-1}{\tilde{\a}}^{\n)}_{-1}|0> & \rightarrow g^{\m \n}\\
          \a^{[\m}_{-1}{\tilde{\a}}^{\n]}_{-1}|0> &  \rightarrow B^{\m\n}\\
          \mbox{Tr}~\a^{\m}_{-1}{\tilde{\a}}^{\n}_{-1}|0> & \rightarrow \phi,
        \end{array}
\right. 
\]
where $g^{\m \n}$, $B^{\m\n}$ and $\phi$ represent the graviton, the 
antisymmetric (Kalb-Ramond) field and the dilaton, respectively. The 
identification of the string states with the quantum fluctuations of 
the corresponding classical fields is justified by two arguments. The first
one takes into account the equivalence under the addition of spurious states 
that is interpreted as gauge transformation. It is easy to see that these
transformations are
\bea
g^{\m \n} & \rightarrow &  g^{\m \n} + \6^\m\xi^\n + \6^\n\xi^\m \\
B^{\m\n} & \rightarrow & B^{\m\n} + \6^\m\xi^\n - \6^\n\xi^\m \\
\phi & \rightarrow & \phi +  \varphi,
\label{gaugemaslesclosed}
\eea 
which are just the gauge transformation of the classical fields. The second
argument relies on the interaction theory and it can be shown that the
states above satisfy the correct equations of motion of the corresponding 
fields.
Thus, we may conclude that the {\em masless spectrum of the closed string 
contains the graviton in D=26}. 

The appearance of gravity in a natural way is
one of the most attractive features of bosonic string theory. However, the
theory suffers from three serious drawbacks: the presence of tachyon, the 
absence of fermions and the high dimensionality of space-time. 
The first two problems can be solved by introducing the
world-sheet supersymmetry and constructing a superstring theory.  It can
be shown that one can obtain a space-time supersymmetric theory  in
the light-cone gauge which is also free of tachyon \cite{gsw,jp}. In the same
time the number of space-time dimensions is reduced from $D=26$ to $D=10$.
The price to be paid is that there are now five consistent superstring 
theories. However, there are strong hints that all these theories are actually 
different limits of an unique underlying theory unknown at present \cite{jp}. 
The other problem, namely reducing the spacetime dimensionality from ten to 
four in a natural (dynamical ?) fashion is unsolved up to day.  

\subsection{Exercises}

{\bf Exercise 1}\\
Starting from the bosonic string action (\ref{polyakov}) find the Neumann and
Dirichlet boundary conditions using Green's theorem in two dimensions:
\be
\int_\Sigma dxdy (\6_x Q - \6_y P) = \int_{\6\Sigma} Pdx + Qdy.
\label{greens}
\ee

{\bf Exercise 2}\\
Prove any of the solutions of the open string with N-N, D-D, N-D and D-N 
boundary conditions and the closed string solution by using the method of 
separation of variables.

{\bf Exercise 3}\\
Starting from equal-time commutator (\ref{eqtcomm1}) show that the
operators $\a$ satisfy the algebra (\ref{alphacomm1}) and that 
the coordinates and the momenta of the center of mass of string satisfy
the relation (\ref{alphacomm2}).

{\bf Exercise 4}\\
Show that the total momentum of the open string defined as
\be
P_{total}^{\m} = \int_C d\s^1 P^{\m}_{0} + d\s^0 P^{\m}_1 ,
\label{conserv}
\ee
where $C$ is an arbitrary curve on the world-sheet from the boundary 
$\s =0$ to $\s = \pi$, is conserved.

\vskip 2cm


\section{Bosonic $D$-branes. Effective Action}

In this lecture I am going to introduce the $D$-branes and to discuss their 
physical degrees of freedom following \cite{jp,cj}. Also, I will present 
the way in which the background field method is applied in order to obtain the
Born-Infeld effective action of D-branes (see \cite{jp,cj,abou}).

\subsection{Definition of bosonic $D$-branes}

By definition, a bosonic $D$-brane is an extended physical object on which 
bosonic open strings can end \cite{jp,cj}. 

As we saw in the previous lecture, the momentum of the open string should be
conserved and it is a reasonable to assume that the momentum does not 
flow away through the ends of the string. However, when the open string ends 
on a $D$-brane, the situation is different. Indeed, there can be an exchange
of momentum between the string and the brane through the end of string that is 
in contact with the brane. Therefore, it is the momentum of the full system
that should be conserved. Another immediate consequence of introducing
branes in string theory is that the Lorentz invariance is now
broken.

The word ``physical'' in 
the definition means that the  branes are characterized by more than their 
geometry and topology and that they have some physical properties like tension
and charge as we shall see in the next lectures. The real motivation for 
introducing the $D$-branes was that in the spectrum of Type I and Type II 
superstring theories there are some bosonic fields which are described by 
$p$-forms in $D=10$ dimensions. At that time it was not known what could have
been the sources of such of fields and it was discovered by Polchinsky that
the $D$-branes were the sought for objects \cite{jpprl}.

It follows from the definition of the $D$-branes that the open strings that
end on them must have Dirichlet boundary conditions on the directions 
transversal to the branes and Neumann boundary conditions on the
directions tangent to the brane. Indeed, there is nothing that can stop the 
string of sliding on the world-volume of the brane. 
In the simplest case a $Dp$-brane is a hypersurface embedded in the $D=10$
space-time where $p$ indicates that it has $p$ space-like directions.
If the $D$-brane is 
situated at the $\s=0$ end of string then the boundary conditions are given
by the following relations
\bea
\mbox{N.b.c.}~ \6_\s X^a|_{\s=0} &=& 0~,~~a=0, 1, 2, \ldots p, \\
\mbox{D. b. c}~~~X^i|_{\s=0} &=& x^i~,~~i=p+1,\ldots ,9.
\label{bcDbrane}
\eea
Actually, by computing the spectrum of the open strings ending on the branes
we can find the degrees of freedom of the brane, that is the fields that live 
on the world-volume. We are interested in the massless degrees of 
freedom which will not change the energy of the brane. They are given by the
strings with both ends on the brane. The other strings will contribute with an
energy proportional to the stretching of the string. For example, the strings
between two branes will contribute with the following stretching 
energy \cite{jp,cj} 
\be 
m^2 = \f{Y^2}{4\pi^2\a '}.
\label{energystretched}
\ee
The relation above can be obtained by considering the T-duality of the theory
\cite{jp,cj} but it can also be established from dimensional arguments. 

In order to find the massless degrees of freedom we take the solution of the
string equations of motion with Dirichlet boundary conditions at the two ends
in the transverse directions (\ref{DDsol}) and Neumann in the tangential 
directions (\ref{NNsol}). After quantizing them as in the previous section we
discover that the massless states are given by
\bea
\a^{a}_{-1} |0>~,~~a &=&0,1,\ldots,p\nonumber\\
\a^{i}_{-1} |0>~,~~i &=&p+1,\ldots,25.
\label{masslessDbr}
\eea
The first set of states describes an $SO(1,p)$ photon $A^a(\xi)$ while the 
second one
is associated to an $SO(25-p)$ massless vector $\Phi^i(\xi)$. 
The components of the latter
are associated to the breaking of the translational symmetry along the 
transverse directions $X^i$ and are interpreted as the fluctuations around
the classical localization of brane in the transverse spacetime. By 
$\xi$ we denoted the coordinates on the world-volume of the $D$-brane.
Actually, the fields $\Phi^i(\xi)$ represent a particular embedding of the 
$D$-brane in spacetime. In general, the corresponding degrees of freedom are
the embedding functions $X^{\m}(\xi)$ of the world-sheet volume in the target 
space. In the above case we have considered the simplest situation in which the
brane was flat and its tangential directions were parallel to some of the
directions of spacetime.

Thus, a $Dp$-brane breaks the space-time symmetry of the theory as follows
\be
SO(1,25) \rightarrow SO(1,p) \times SO(25-p)
\label{breaklorentz}
\ee
and consequently its massless degrees of freedom are given by the set
$\{ A^a(\xi), X^{\mu}(\xi) \}$. 

\subsection{Effective action of $Dp$-branes}

It is possible now to find the dynamics of the $Dp$-brane in the low energy
limit. Indeed, in this limit the degrees of freedom of the brane are the 
classical fields found in the previous section. One should look for an
action describing the dynamics of these fields and we end up with the effective
field theory of the brane.

Recall that the degrees of freedom of the branes were found in terms of 
open strings ending on them. In order to have a
description consistent with two dimensional string theory, one 
should stick on the conformal invariance of string in the new 
background in which the $Dp$-branes are present (\cite{jp,cj}). Without the 
conformal invariance the two dimensional theory will not describe 
a physical theory. This requirement is the same for strings in {\em any 
arbitrary} background and it is implemented as follows. 
The string theory in a general background contains couplings between strings
and the background fields. However, these coupling terms break in general the
conformal invariance. In order to find those configurations which preserve the
two dimensional conformal invariance, one treats the background fields as 
coupling constants and the sought for configurations can be found by solving
the equation
\be
\b_{\Gamma} = 0,
\label{beta1}
\ee
where $\b_{\G}$ is the beta-function of any background field $\G$ 
\cite{gsw,jp}. 
One way to compute the beta-functions is by using the {\em background field 
method} (see \cite{ag,at}.) We are going to show how this method is
applied to obtain the low energy action of $Dp$-branes following 
(\cite{callan}).

In a background containing $Dp$-branes the open strings couple with the 
brane degrees of freedom $\{ A^a(\xi), X^{\mu}(\xi) \}$. Beside them, there
may be other massless fields in the background like, for example, the closed
string fields $g_{\m \n}, B_{\m \n}, \phi$. If the theory is supersymmetric,
then massless fermions are also present. Each of these background fields will
have a beta-function that must vanish if the two-dimensional theory that 
describes strings is to be conformal. 

\vskip 0.5cm
{\bf $D25$-brane in a flat background}
\vskip 0.5cm

To understand how the beta-functions are computed, let us consider firstly a
simpler situation in which we have a $D25$-brane that fills the whole 
space-time and no closed string fields in the background. The only
background fields are $A^{\m}(X)$. The photon couples with one dimensional 
world-line or, equivalently, with dimensionless charges like in 
electrodynamics. Therefore, in order to couple it with the string which has
a two-dimensional world-sheet, we have to put some ``electric charges'' at 
the ends of the open strings (which are just points) and to couple the photon
with these charges in the usual way. Actually, this explains the presence
of the $U(1)$ field on the world-volume of the brane as being generated by 
the charges at the end of the open string.  
Let us consider both the space-time and the world-sheet Euclidean and map 
the world-sheet to the
complex upper-half plane with $z=\t + i\s$. The $U(1)$ field couples on the 
boundary of the world sheet and the total action is given by the formula
\be
S = \f{1}{4 \pi \a '}\int_{\Sigma} d^2 z \6_\a X^\m \6^\a X_\m +
\f{i}{2\pi \a '}\int_{\6\Sigma}d\t A_{\m}(X)\dot{X}^\m,
\label{totalactinter}
\ee
where $A_\m$ has been rescaled to include a $2\p\a '$ factor
and the $U(1)$ charge has been taken 1.  Choose a 
background field $\bar{X}^\m(\t,\s)$ that is a solution of the 
equations of motion 
and of boundary conditions that are derived from Eq.(\ref{totalactinter}) 
above. Now expand the fields $X(\t,\s)$ arround this solution. One obtains the
following set of equations
\bea
X^\m(\t,\s) = \bar{X}^\m(\t,\s) &+& \z^\m (\t,\s)\nonumber\\
\Box \bar{X}^\m(\t,\s) & = & 0\nonumber\\ 
\6_\s \bar{X}^\m + iF^{\m}_{\n}\6_\t \bar{X}^{\n}|_{\6 \Sigma} &= &0,
\label{backgroundall}
\eea
where $\Box = \6^{2}_{\s} + \6^{2}_{\t}$ is the Laplacean on the Euclidean
world-sheet, $F_{\m\n} = \nabla_{[\m}A_{\n]}$ and $\nabla_\m = \6/\6X^\m$.
The full information on the field is contained in the fluctuation $\z$ around
the background $\bar{X}$ \cite{ag,at,call}. Introducing (\ref{backgroundall})
in the action (\ref{totalactinter}) we obtain the expansion arround the
background solution. We consider only slow varying fields $F_{\m\n}$. This
condition allows us to disregard higher derivatives of $F$ like $\nabla^2 F$, 
$\nabla^3 F$, \ldots . The expanded action takes the form
\bea
S[\bar{X} + \z] &=& S[\bar{X}] + \f{1}{2 \pi \a '}\int_{\Sigma} d^2 z
(\6_\a {\bar{X}}^\m \6^\a \z_\m + \f{1}{2}\6_\a \z^\m \6^\a \z_\m + \cdots) \nonumber\\
&+& \f{i}{2 \pi \a '}\int_{\6 \Sigma}d\t(F_{\m\n}\z^\m \6_\t \bar{X}^\n
+ \f{1}{2}\nabla_\r F_{\m\n}\z^\r \z^\n \6_{\t} \bar{X}^\m \nonumber\\
&+& \f{1}{2}F_{\m\n}\z^\m \6_\t \z^\n + 
\f{1}{3}\nabla_\r F_{\m\n}\z^\r \z^\n \6_\t \z^\m + \cdots).
\label{actionexpand}
\eea
We look for the one loop beta-function. This is given by the one-loop
counterterm with one external leg $\6_\t \bar{X}$ of the interaction term
in (\ref{totalactinter}), that is
\be
i\int_{\6\sigma}d \t A_\m\6_\t \bar{X}^\m \rightarrow
\triangle S_I[\bar{X}] = \f{i}{2 \pi \a '}\int_{\6\Sigma}d\t \G_\m \bar{X}^\m .
\label{counterone}
\ee 
The value of the corresponding one-loop Feynman diagram is
\be
\triangle S_I[\bar{X}] = -\f{i}{2 \pi \a '}\int d\t \f{1}{2}\nabla_\r
F_{\m\n}\6_\t \bar{X}^\m G^{\r\n}(\t,\t ')|_{\t\rightarrow\t '},
\label{countertwo}
\ee
where $G$ is the Green's function computed on the boundary $\s =0$ in the 
point $\t$. Thus, it is the solution of the following problem
\bea
\f{1}{2\pi \a '}\Box G_{\m\n}(z,z') &=& -\delta_{\m\n}\delta(z-z')\\
\6_\s G_{\m\n} &+& iF^{\l}_{\m}\6_\t G_{\n\l}|_{\6\Sigma} =0.
\label{greenproblem}
\eea
One can find the explicit form of the Green's function by using the
method of images \cite{call} and it is given by the following relation
\be
G_{\m\n} = \a '[\d_{\m\n}\ln |z-z'| + \f{1}{2}(\f{1-F}{1+F})_{\m\n}
\ln (z-\bar{z}') +  \f{1}{2}(\f{1+F}{1-F})_{\m\n} \ln (\bar{z}-z')], 
\label{greensolution}
\ee
where we have used the notation
\be
(\f{A}{B})_{\m\n} = A_{\m}^{\r}(B^{-1})_{\r\n}.
\label{matrixproduct}
\ee
When $F=0$ Eq.(\ref{greensolution}) reduces to the known Green's function 
in the absence of
$U(1)$ field. The Green's function on the boundary is 
\be
G_{\m\n}(\t\rightarrow\t ') = -2\a '\ln \L (1-F)^{-1}_{\m\n},
\label{greenboundary}
\ee
where $\L$ is a short distance cut-off. The beta-function of the field
$A^\m$ is given by applying the definition and it should vanish in order to 
have a conformal invariant theory
\be
\b^{A}_\m = \L \f{\6\G_\m}{\6\L} = \nabla^\r F^{\l}_{\m}(1-F^2)^{-1}_{\l\r} =0.
\label{betaA}
\ee

The effective action is the action from which the Eq.(\ref{betaA}) can be 
defined through the variational principle, i. e. the action which has 
the equations of motion given by Eq.(\ref{betaA}). Actually, there is no
such of action \cite{call} and therefore one has to find an equation that is
equivalent to (\ref{betaA}). Such of equation is 
\be
\chi^{\m\n}(F)\b^{A}_{\n} = 0,
\label{equiveqA}
\ee
for any invertible matrix $\chi^{\m\n}(F)$. Now after some algebra 
\cite{callan}, one can show that the sought for equation of motion
is 
\be
\sqrt{\mbox{det}~(1+F)}(1-F^2)^{-1}_{\m\n}\b^{\n}_{A}=0,
\label{fineq}
\ee
which can be derived from a non-polynomial action called 
{\em Born-Infeld action} given by the integral of
\be
L_{BI}= \exp [\f{1}{2}\mbox{Tr}\ln (1+F)] = [\mbox{det}(1+F)]^{\f{1}{2}}.
\label{borninfelone}
\ee 
The above Lagrangian describes the effective action of the massless states of 
the open string in a background that contains an $U(1)$ gauge potential that
couples with the boundary of the world-sheet. According to D-brane 
interpretation this is the effective action of a $D25$-brane. 

\vskip 0.5cm
{\bf $D25$-brane in closed string background}
\vskip 0.5cm

The situation can be complicated further to include other fields in background.
Since all the fields should come from the string spectrum, we may include
other massless or massive string fields. Let us consider a background in 
which the graviton, the 
dilaton and the Kalb-Ramond two-form field do not vanish. The action 
for the open string contains the following terms
\be
S= S_g + S_B + S_\phi + S_A
\label{generalaction}
\ee
where
\bea
S_g &=& \f{1}{4\pi\a '}\int_{\Sigma}d^2 z g_{\m\n}(X)\6_\a X^\m \6^\a X^\n
\nonumber\\
S_B &=& - \f{i}{4\pi\a '}\int_{\Sigma}d^2 z 
\e^{\a\b}B_{\m\n}(X)\6_\a X^\m\6_{\b}X^\n
\nonumber\\
S_\phi &=& \f{1}{4\pi\a '}\int_{\Sigma}d^2 z (-\f{1}{2}\a ')\sqrt{h}
R^{(2)}\phi(X) + 
\f{1}{2\pi\a '}\int_{\6\Sigma}d\t (-\f{1}{2}\a ')k\phi(X)
\nonumber\\
S_A &=& 
\f{i}{2\pi\a '}\int_{\6\Sigma}d^2 z A_{\m}(X)\6_\t X^{\m},
\label{termsinaction}
\eea
where $\e_{\a\b}$ is the two-dimensional antisymmetric symbol, $h$ is the 
determinant of the two-dimensional metric $h_{\a\b}$ which at the tree-level is
Minkowski, at one-loop is cylindric, etc. and $R^{(2)}$ is the 
two-dimensional curvature. 

The background field $A^\m$ is  treated as a coupling constant as in the 
previous example. In order to have a conformal invariant field theory in
two-dimensions, the beta-function of it should vanish. The 
beta-function can be computed using the same method as above. The 
field $X^\m$ is chosen to satisfy the free field theory with interaction 
with the 
Kalb-Ramond and $U(1)$ gauge-potential on the boundary. The world-sheet is the
same as in the previous case. The equations of motion and the boundary
conditions that are obtained from the sigma-model (\ref{generalaction}) are
given by the following relations (\cite{at,call})
\bea
[g^{\m}_{\n}\6_\a &+& \G^{\m}_{\n\l}\6_\a\bar{X}^{\l} +
\f{i}{2}H^{\m}_{\n\l}\e^{\a\b}\6_\a\bar{X}^\l]\6_\a\bar{X}^{\n}=0
\\
\6_n \bar{X}^\m &-& i(B+F)^{\m}_{\n}\6_\t\bar{X}^{\n}|_{\s = 0}=0,
\label{eqmotboundcondsigma}
\eea
where
\be
H_{\m\n\r}= 3\nabla_{[\r}B_{\m\n]}.
\label{hash}
\ee
The contribution to the compensating term, at one-loop, gives the following  
beta-function of $A^{\m}$ \cite{call}
\bea
\b^{A}_{\m} &=& \nabla^{\r}(B+F)^{\m}_{\n}[g-(B+F)^2]^{-1}_{\n\r}
\nonumber\\
 &+&
\f{1}{2}(B + F)_{\m\n}H^{\n\l\r}[\f{B+F}{g-(B+F)^2}]_{\l\r}+
\f{1}{2}\nabla^{\n}\phi(B+F)_{\m\n}
\label{betaAsigma}
\eea
The invertible matrix that generates a variational equation from the
equations that imposes the conformal invariance on the system is
\be
\chi_{\m\n} = (g-(B+F)^2)^{-1}_{\m\n}
\label{chieq}.
\ee
The Born-Infeld effective action is given by the following relation
\be
S_{eff} \simeq \int d^{26}X e^{\f{1}{2}\phi} [\mbox{det}~(g + B + F )]^{1/2}
\label{effactsigma}
\ee
from which one obtains the following equation of motion
\be
e^{-\f{1}{2}}[\mbox{det}(g+B+F)]^{\f{1}{2}}(g-(B+F)^2)^{-1}_{\m\n}\b^{\n}_A =0.
\label{eqiveqmotioneff}
\ee
The $\simeq$ means that the action is given up to some dimensional constant.
This constant is necessary in order to make the left hand side of 
(\ref{effactsigma}) an action and from dimensional arguments one sees that it 
should have the dimension of the brane tension $T$.

\vskip 0.5cm
{\bf $Dp$-brane effective action}
\vskip 0.5cm

The Born-Infeld action of the low energy effective field theory of a generic
$Dp$-brane can be obtained as in the two examples above. Actually, one can 
easily adapt the action (\ref{effactsigma}) to serve this purpose. To this
end, note that in the case of the $Dp$-branes the fields will interact with 
the $(p+1)$-dimensional world-volume. As was discussed at the beginning of this 
section, $A^\m$ are fields living on the world-volume, therefore they will 
depend on the world-volume coordinates $\xi$. The rest of the fields live in
the full space-time, but they interact with the world-volume through some
``world-volume projected'' components. This projection is given by the 
pull-back of the embedding $X^{\m}(\xi)$ of the world-volume into the
space-time. We denote by $\hat{}$  these fields. Then the Born-Infeld action is
given by
\be
S_{Dp} = -T_p \int \d^{p+1}\xi e^{-\hat{\phi}}[\hat{g}_{ab} + \hat{B}_{ab}
 +  2\pi \a ' F_{ab}]^{1/2},
\label{borninfeld}
\ee
where $T_p$ is the tension of the brane, the pull-back of the fields are
\bea
\hat{g}_{ab}& = & \6_a X^\m\6_b X^\n g_{\m\n}\nonumber\\
\hat{B}_{ab} & =& \6_a X^\m \6_b X^\n B_{\m\n}\nonumber\\
\hat{\phi} &=& \phi(\xi),
\label{pullbacked}
\eea
and $\6_a$ denotes the world-volume derivative $\6/\6\xi^a$. The field-strength
of the $U(1)$ form is a world-volume field and therefore is not pulled-back.
The dilaton coupling is $-1$ since we consider the coupling with the disk.

The action (\ref{borninfeld}) has two gauge symmetries, a $U(1)$ one and a
Kalb-Ramond one, given by the following transformation laws
\bea
\delta A &=& d\lambda ~~;~~\delta B = 0\\
\delta B &=& \d\zeta ~~;~~\delta A = -\f{1}{2\pi \a '}\zeta .
\label{gaugetr}
\eea
Only the combination 
\be
\hat{B}_{ab} + 2\pi \a 'F_{ab}
\label{combination}
\ee 
is invariant under both gauge transformations, which explains the presence of 
the pull-back of the Kalb-Ramond field in the action, even if it does not
couple directly with the world-volume of the $Dp$-brane. 
The factor 
\be
\e^{-\phi}=g^{-1}_{s}
\label{stringcoupling}
\ee
is proportional with the inverse of the string coupling. Therefore, by 
varying the dilaton expectation value, one can study the dynamics of $D$-branes
in different regimes. This situation is familiar from string theory in which
the coupling constant is dynamical. If we take
$F_{ab}=\hat{B}_{ab}=\phi=0$ then the action is proportional to the geometric 
volume of the world-volume
\be
S = -T_p \int d^{p+1}\xi\sqrt{\hat{g}}.
\label{volume}
\ee

\subsection{Exercises}

{\bf Exercise 1}\\
Calculate the expansion in (\ref{actionexpand}).

{\bf Exercise 2}\\
Find the Green's function for the two dimensional Laplace operator from
\be
S = \f{g}{4\pi}\int d^2 z \6\Phi\bar{\6}\Phi
\label{green2d}
\ee
and put the appropriate boundary conditions.

{\bf Exercise 3}\\
Find the Green's function for the following problem
\bea
\f{1}{2\pi\a '} \Box G(z,z') & = & -\delta(z-z')\nonumber\\
\6_\s G(z,z')|_{\6\Sigma} &=& 0
\label{green2done}
\eea
on the upper half-plane ($z=\t + i\s$).

{\bf Exercise 4}\\
Using the Fourier transformation of the Green's function in the upper 
half-plane
\be
G(z,z') = \int \f{dp}{2\p}\f{e^{ip(\t -\t ')}}{2|p|}
[e^{-|p||\s - \s '|} + e^{-|p|(\s + \s ')}]
\label{greenfourier}
\ee
find $G_{\m\n}(\t \rightarrow \t ')$ on the boundary.

{\bf Exercise 5}\\
Prove the following identity
\be
(1-F^2)^{-1}_{\m\n}\b^{\n}_{A} =
\nabla^{\n}(\f{F}{1-F^2})_{\m\n} - (\f{F}{1-F^2})_{\m\l}\nabla^{\n}F^{\l\r}
(\f{F}{1-F^2})_{\r\n},
\label{identity}
\ee
where
\be
\b^{\n}_{A}=\nabla^\n F^{\l}_{\m}(1-F^2)^{-1}_{\l\n}.
\label{betafin}
\ee 

\vskip 2cm


\section{Bosonic D-branes. Tension}

In this section we are going to compute the tension of the D-branes by
computing the interaction amplitude in the string theory and then 
comparing it with the field theory computations.

\subsection{String computation}

The D-branes interact by exchanging closed strings in various quantum states
in analogy with the interaction between particles that exchange some other
(virtual) particles. There is some response in the brane to the exchange of 
closed string excitations and this response should be proportional to the tension 
of the brane. The quantity that measure the intensity of the exchange of 
closed string states is the exchange amplitude.

We are going to do this computation at the tree level in perturbation string
theory because we want to compare latter the result with the corresponding 
calculations in the low energy limit field theory. Since in this limit only 
massless quanta participate to the interaction, we have to take into account
only the effects produced by these string modes. 

One way to do these computations is to interpret the tree level Feynman diagram
for closed strings as one-loop diagram for open strings. Let us see how this 
is done. At tree level, a closed string emitted at the moment $\t =0$ 
propagates along the cylinder an interval $T$. Therefore, the horizontal
coordinate of the cylinder is $0 \leq \t \leq T$ and the periodic one is 
$ 0 \leq \s \leq \pi$, the space-like parameter of the closed string.
However, the same cylinder can be interpreted as an open string of
length $0 \leq \s \leq \pi$ that propagates on a loop in the time $0\leq \t
\leq \hat{T}$. In this case the horizontal coordinate of
the cylinder is parametrized by $\s$. Then the two amplitudes in closed string
and open string description (called also {\em channels}) should be equal. To
have the same cylinder in the two cases, the parameters $(\t,\s)$ of the
closed and open strings should be adjusted in such of way that the interval
$T$ parametrized by $\t$ in closed string channel be equal to $\pi$ 
parametrized by $\s$ in open string channel, which gives $\hat{T}=\pi / T$. 

\vskip 0.5cm
{\bf The amplitude in the open string channel}
\vskip 0.5cm

The one-loop vacuum amplitude in QED is given by the logarithm of the partition
function 
\be
{\cal A}=\ln (Z_{vac})
\label{qftpart}
\ee
and it can be calculated by using the Coleman-Weinberg formula that can be 
obtained as follows \cite{cj}. Start with 
the logarithm of the partition function for a 
scalar field given by the following relation
\cite{jp,cj} 
\be
\ln (Z_{vac})=-\frac {1}{2} \mbox{Tr} \ln ({\Box}+ m^{2})= - 
\frac {V_{d}}{2} \int \frac {d^{d}k}{(2 \pi)^{d}} \mbox{Tr} \ln (k^2 + m^2),
\label{lnscal}
\ee
where  $d$ is the number of the dimensions of space-time, $V_d$ is the 
volume in which the field is contained and $ (k^2 + m^2)/2=H$ is the 
Hamiltonian of the field. Then use the following 
following property of the $\ln$ function
\be
\ln x= -\lim_{\e \rightarrow 0} \int^{\infty}_{\e} \frac {dt}{t} e^{-t x}
\label{lnprop}.
\ee
By inserting (\ref{lnprop}) into (\ref{lnscal}) one obtaines the 
Coleman-Weinberg formula for a scalar field 
\be
{\cal A} = V_{d} \int \frac {d^{d}k}{(2 \pi)^{d}}\int^{\infty}_{0} 
\frac {dt}{2t} 
\mbox{Tr}~e^{-(k^2 + m^2)t/2}.
\label{ColemanWeinberg}
\ee
When one integrates on the circle, the two orientation of it are taken into 
account, that is why we have to divide the integrand of 
(\ref{ColemanWeinberg}) by a factor of 2. 
The relation (\ref{ColemanWeinberg}) has the interpretation of the 
{\em free energy}.

We are going  to apply now the formula (\ref{ColemanWeinberg}) to the modes 
of the open strings
that move on the circle parametrized by $ 2\pi \t$ at one-loop. We know from 
the second section that the Hamiltonian of the open string is given by 
\be
H=L_{0}-1=\a^{'}(k^2 + {\hat{M}}^{2})
\label{hamiltopen}
\ee
where the Virasoro operator $L_0$ is given by the relation
\be
L_{0}=\a^{'}k^2 + \a^{'} \frac {Y^{2}}{(2 \pi \a^{'})^{2}} + 
\sum^{\infty}_{n=1} \a_{-n} \cdot \a_{n}.
\label{virzero}
\ee
The term proportional to the distance $ Y^2 $ between the branes is due to
the stretching energy of the string. Then the amplitude for the open string 
modes can be computed by applying directly the formula (\ref{ColemanWeinberg}).
This is possible since the string can be viewed as a collection of scalar 
fields in two-dimensions. Then the sought for amplitude is given by the 
following relation
\be
{\cal A}= \int^{\infty}_{0} \frac {d \t}{2 \t} 
\mbox{Tr}~e^{ -2\pi \t(L_{0} - 
1)}. 
\label{amplitopenstring1}
\ee
or, by plugging (\ref{virzero}) into (\ref{amplitopenstring1}) 
\be
{\cal A} = \int^{\infty}_{0} \frac {d \t}{\t}  
\times V_{p+1} \int \frac{d^{p+1}k}{(2 \pi)^{p+1}} 
e^{-2\pi \t \a^{'}k^{2}} e^{-\frac {y^{2} \t}{2 \pi \a^{'}}}e^{2 \pi \t} 
\mbox{Tr}~\left[ e^{-2\pi \t \sum^{\infty}_{n=1} \a_{-n} \cdot \a_{n} }\right].
\label{amplitopenstring2}
\ee
The factor of 2 disappeared from the denominator since we allow the interchange
of the orientation of open string and each orientation gives and equal
contribution to the amplitude.
Note in (\ref{amplitopenstring1}) the expression of the mass-shell condition 
$ ( {\Box} + m^{2}) \phi=0  \leftrightarrow (L_{0} - 1) |\phi>=0$ in the QFT 
and string theory, respectively. From the last relation we will write the 
propagator of the string later.

To compute the r.h.s. of (\ref{amplitopenstring2}) we note that it factorizes
into an integral over $k$ and the trace over the oscillation modes. The 
integral is Gaussian and from it we will obtain the factor 
\be
(8\pi^2\a '\t)^{-\f{p+1}{2}}.
\label{factor}
\ee
Also, by computing the trace of $\hat{k'}^2$ in the parallel directions to the
world-volume of the brane, the volume $V_{p+1}$ of the brane is obtained. We 
use the following normalization relation
\bea
<k|k^{'}>&=&2 \pi \d (k-k^{'}) \nonumber \\
V_{p+1}&=& (2 \pi)^{p+1} \d ^{p+1}(0).
\eea

The trace over the oscillators can be computed in the basis of the operators
$\a_{-n}$ and $ \a_n$ 
\bea
\mbox{Tr} e^{-2\pi \t \sum^{\infty}_{n=1} \a_{-n} \cdot \a_{n}} &=& 
\prod ^{\infty}_{n=1} \prod^{25}_{\mu=0} \mbox{Tr}~e^{-2 \pi \t 
\a_{-n}^{\mu}\a_{n \mu}}
=\prod ^{\infty}_{n=1} \prod^{25}_{\mu=0} \sum^{\infty}_{m=0} <m|e^{-2 \pi
\t n a_{-n}^{\dag} a_{n}}|m>\nonumber\\
&=&\prod^{\infty}_{n=1} \left ( 
\frac {1}{1- e^{-2\pi \t n}} \right )^{26}
\label{traceosc}
\eea
where $a_{n}^{\dag}a_{n}|m> = m|m>$. The trace above includes the contribution 
of the non-physical degrees of freedom. To remove them, one should pick-up
a gauge. In any covariant gauge, the non-physical degrees of freedom are 
taken account of by the Fadeev-Popov ghosts. Without writing their explicit
contribution we give the final form of the amplitude
\be
{\cal A} =V_{p+1}(8 \pi^{2}\a^{'})^{- \frac{p+1}{2}} \int ^{\infty}_{0} 
\frac{d \t}{\t} \t^{- \frac{p+1}{2}} e^{-\frac {Y^{2} \t}{2 \pi \a^{'}}} 
\left [f_{1}(e^{-\pi \t}) \right ]^{-24}.
\label{finalamplitopen}
\ee
We see that the effect of the ghosts is to reduce the number of space-time
dimensions by two, i.e. to the transverse directions. This can be also done 
by solving firstly the constraints, which will leave the theory in the 
light-cone gauge \cite{gsw}. The function $f_1$ is defined as
\be
f_{1}(q)=q^{\frac {1}{12}} \prod ^{\infty}_{n=1} (1-q^{2n})
\label{f1}
\ee
and under a {\em modular transformation} of its variable
\be
\t \longrightarrow \frac {1}{\t}
\label{modulartrt}
\ee
it transforms in the following way
\be
f_{1}(e^{-\pi/ \t})= \sqrt{\t}f_{1}(e^{-\pi \t})
\label{modularf1}
\ee
which is the modular transformation property of the function $f_1(q)$.

\vskip 0.5cm
{\bf The amplitude of closed string massless modes}
\vskip 0.5cm

Now we would like to identify in the amplitude (\ref{finalamplitopen}) the
contribution of the closed string modes which interest us. To this end, we 
note that in the limit $\t \rightarrow \infty$, i.e. when the circle of
the cylinder opens, the world-sheet becomes a long and thin strip. In
order for a mode of the open string to travel the loop, it should be light
since it has to reach across a long distance. Thus, in this limit the
light modes of the open string dominate the amplitude. In the limit 
$\t \rightarrow 0$, the open string is in the UV regime since the radius of
the circle is small and the string modes have to travel short distances in 
making the loop. However, this limit is the long-distance of the closed string.
Indeed, by making a reparametrization of the string length (a conformal
transformation) that does not change the area of the cylinder while it makes it
radius small, we see that the length of the cylinder goes as
\be
Y_{1}=\frac {Y_{0}}{2\e } \longrightarrow \infty ,
\label{lengthcyl}
\ee
for any scale unit $\e = \t \rightarrow 0$, where $Y_0$ is the distance 
between branes. $Y_1$ is the apparent length of the cylinder as viewed by the
string modes. In the closed string channel, the closed string modes have to 
travel this distance between branes and therefore this is the UV limit of
closed strings in which its light modes have a major contribution.
 
All we have to do now is to make the modular transformation of the
cylinder parameter $\t$ given by (\ref{modulartrt}) and to make the
expansion of $f_1$ function in $\t \rightarrow 0$ limit 
\be
\left [f_{1}(e^{-\pi x}) \right ]^{24}_{x \rightarrow \infty}= 
\sum_{n=0} c_{n} e^{-2 \pi x(n-1)}= e^{2 \pi x} +24 + O(e^{-2 \pi x}).
\label{powerxpanf1}
\ee
Each term in the power expansion corresponds to the trace of closed string 
modes with mass
\be
\frac {\a^{'}M^{2}}{2}=2(n-1).
\label{closedmass}
\ee
The first term is the contribution of the tachyon and we are only
interested in the second term which represents the contribution of the 
closed string massless modes. The sought for interaction amplitude is 
given by the following relation
\be
{\cal A}=V_{p+1}\frac {24 \pi}{2^{10}}(4 \pi^{2} \a^{'} )^{11-p} G_{25-p}(Y)
\label{closedstringfinampl}
\ee
where the Green's function in the transverse directions to the world-volume 
of the brane is given by the relation
\be
G_{25-p}(Y)=2^{-2} \pi^{\frac {25-p}{2}} \G (\frac{1}{2}(25-p)-1)Y^{2+p-25}.
\label{green}
\ee
Here, $\G \left( \frac{1}{2}(25-p)-1 \right)$ is the Gamma-function.

\subsection{ Field theory computations}

The amplitude that has been obtained in (\ref{closedstringfinampl}) describes
the interaction of the $Dp$-branes via the exchange of closed string
massless modes. We saw in the second section that these modes are identified
with the quanta of the gravitational, dilaton and Kalb-Ramond fields. We saw
in the previous section that the low energy effective field theory that
describes the $Dp$-brane dynamics is given by a Born-Infeld action. This
action was obtained by requiring the conformal invariance of the open strings
in the background that contains a $Dp$-brane.

This idea can be applied to string field in any background. The result will
be, as in the case of the $Dp$-brane, an effective action that describes the
dynamics of the background fields. Not all background fields will preserve
the conformal invariance of the string theory but only those ones that 
satisfy the vanishing beta-function condition which is identified with the
equations of motion of the classical fields. 

In a general background that contains only $\phi$, $G_{\m\n}$ and $B_{\m\n}$
fields, the effective action of them is given by the following $\s$-model
action  
\be
S= \frac{1}{2k_{0}^{2}} \int d^{26}X \sqrt{-G} e^{-2\phi} 
\left [R +4D_{\mu}\phi D^{\mu}\phi - \frac{1}{12}H_{\mu \nu \rho}H^{\mu \nu \rho} 
\right] ,
\label{closedeffectaction}
\ee
where $H$ is the field-strength of $B$
$$
H_{\mu \nu \rho}=\6 _{[\m}B_{\n \r ]}
$$
and $D_\m$ is the space-time covariant derivative \cite{gsw,jp}. The action
(\ref{closedeffectaction}) can be obtained by using the background field
method exposed in the previous section. Its equations of motion are equivalent
to the conformal condition
\be
\b^{\phi} = \b^{G}_{_\m\n} = \b^{B}_{\m\n} = 0,
\label{backconf}
\ee 
where all the fields are treated as coupling constants and open string fields
are absent from the background. 

The action (\ref{closedeffectaction}) describes the dynamics of the fields
we are interested in in the bulk. However, before computing the interaction 
amplitude from it, we would like to decouple the dilaton from the curvature
in order to benefit from the results of general relativity. The 
effective action in (\ref{closedeffectaction}) is known as being in the
{\em string frame}, and we want to write it in the {\em Einstein frame} by
making the following rescaling of the metric 
\be
g_{\m \n}= e^{\frac{\phi _{0}-\phi}{6}}G_{\m \n},
\label{passeinstein}
\ee 
where $\phi_0$ is the v.e.v. of the dilaton.

In order to find the quantum amplitude we go to the linearized form of the 
action (\ref{closedeffectaction}). To this end, we expand the the background
field around their classical values
\bea
\phi &=& \phi_{0} + \k _{0} \nonumber\\
g_{\mu \nu} &=& \eta_{\m \n }+ \k_{0}h_{\m \n }
\label{expandvalue}
\eea
where $k_0$ is the gravitational coupling constant and the expansion 
parameter. Then we set all the v.e.v. to zero with the exception of 
$\eta_{\m\n}$.

In order to construct the Feynman diagrams that describe the interaction
between fields and branes we must know the coupling constants between these
fields and the $Dp$-branes. The coupling constant are given by the interaction
term which is the linearized form of the effective action of the brane. We
recall that in the string frame it is given by the relation
\be
S_{BI}=-T_{p} \int d^{p+1}\xi e^\phi \le [\mbox{- det}(\hat{G}_{ab}+
\hat{B}_{ab} + 2 \pi \a^{'}F_{ab} \rh) ]^{1/2}.
\label{borninfeldint}
\ee
To the leading order in the the gravitational coupling, the interaction between
fields and branes has the form 
\be
\le [\mbox{-det}~(\h_{ab} + k_{0}(h_{ab}+B{ab}+2\pi \a^{'}F_{ab})) \rh ]^{1/2}
= \frac{3}{2} + \frac{1}{2} \k_{0} h^{a}_{a} + O(\k_{0}^2).
\label{leadingint}
\ee 
We see that the antisymmetric tensors decouple and the only contribution
is from dilaton and graviton. Then the linearized action of the fields that
interact with the $Dp$-brane, in the Einstein frame, have the following form
\be
S_{int}= \frac{1}{2k_{0}^{2}} \int d^{26}X \sqrt{-g} \left (R -\frac{1}{6} 
D_{\mu} \phi D^{\mu}\phi \right)
\label{bulklinear}
\ee
for the bulk action and
\bea
S_{BI~int}&=&- \frac{T_{p}}{\k_{o}} \int d^{p+1}\xi 
e^{\frac {p-11}{2}\phi} \sqrt{\mbox {-det}\hat{g}_{ab}}.
\nonumber\\
&=& S_{cl} - T_{p} \int d^{p+1} \xi \le (\frac{p-11}{2} \phi + 
\frac {1}{2} h^{a}_{a} \rh)+ O(\k_{0}),
\label{BIlinear}
\eea
where $S_{cl}$ is the classical action. Since we are interested in the 
classical effects we are going to compute the tree level amplitude. 
The brane acts as sources of fields 
which propagate from one brane to the other. Now let us compute the 
propagators.

The linearized part of the graviton interaction is given by the following 
Lagrangian
\be
{\cal L}_{int}= -\frac{1}{2} \6_{\l } h^{\l }_{\m } \6^{\m } h^{\n }_{\n } +
\frac{1}{2} \6_{\l} h^{\l }_{\m } \6_{\n } h^{\n \m } - 
\frac{1}{4}\6_{\l }h_{\m \n  } \6^{\l } h^{\m \n }+ 
\frac{1}{4}\6_{\l } h_{\m }^{\m } \6^{\l }h_{\n }^{\n }
\label{gravactprop}
\ee
with the gauge gauge invariance
\be
h_{\m \n } \longrightarrow h_{\m \n } +\6_{\mu}\xi_{\nu} + \6_{\nu}\xi_{\mu}.
\label{gaugegraviton}   
\ee
Since there are gauge degrees of freedom, one has to fix the gauge by adding 
a gauge breaking term which can be chosen to be
\be
{\cal L}_{c}= - \frac{1}{2} C^{2},
\label{gaugebreak}
\ee
where
\be
C^{\mu}=\6_{\nu}h^{~ \mu}_{~ \nu} - 
\frac{1}{2}\6^{\m }h^{\n }_{~\n}. 
\label{constraint}
\ee
The gauge fixed Lagrangian is the sum between (\ref{gravactprop}) and 
(\ref{gaugebreak})
\be
{\cal L}_{int~gauge~fixed}={\cal L}_{int}+{\cal L}_{c}= - 
\frac{1}{2}[\6_{\l}h_{\m \n}\6^{\l }h^{\mu \nu} - 
\frac{1}{2}\6_{\l}h_{\mu}^{~ \mu} \6^{\l }h^{~\n}_{ \nu }]. 
\label{gravitact}
\ee
By integrating by parts to put in evidence the propagator and adding
the dilaton part we obtain the following relation
\be
S_{int}= -\frac{1}{2\k_{0}} \int d^{26}X \le \{ \frac{1}{2} h_{\mu \nu} 
\left [\h^{\m \l }\h^{\nu \s}+\h^{\m \s }\h^{\nu \l }- 
\frac{1}{12}\h^{\m \n }\h^{\l \s} \rh ] \6^{2}h_{\l \s } + \frac{1}{6} 
\phi \6^{2}\phi \rh \}.
\label{gaugefixedgravdil}
\ee
By definition, the propagators are given by the functional derivatives of
the action with respect to the fields
\bea
D_{\m \n , \l \s } &=& - \frac{\d^{2} S_{int}}{\d h^{\m \n } 
\d h^{\l \s}} \le .\rh |_{h_{\m \n }=0}
\label{gravpropag}\\
D &=& - \frac {\d^{2} S_{\mbox{int}}}{\d \phi  
\d h^ \phi {'}} \le . \rh |_{\phi=0}
\label{dilatpropag}
\eea 
It is easy to write down the propagators for the graviton and the dilaton
\bea
D_{\m \n , \l \s }(\k) &=& -2 \k^{2}_{0}  \left (\h^{\m \l }\h^{\nu \s} +
\h^{\m \s }\h^{\nu \l } - \frac{1}{12} \h^{\m \n }\h^{\t \s} \rh ) 
\frac{1}{\k^{2}},
\label{gravprop}\\
D(\k ) &=& -6 \k^{2}_{0} \frac {1}{\k^{2}}.
\label{propagdilat}
\eea
The  currents  necessary to write down the values of the Feynaman 
diagrams can be read off the action linearized action (\ref{BIlinear}) and 
are simply the coefficients of the fields 
\bea
j_{\phi} &=& \frac{p-11}{12}T_{p} \d_{\bot}
\label{currdil}\\
T_{\m \n} &=& \frac{1}{2} T_{p} \d_{\bot} \times 
\left\{ \begin{array}{cc}
    \eta_{\m\n} & \mbox{for} \m ,\n \neq p\\
    0           & \mbox{in the rest}\nonumber
    \end{array}
\right.
\label{currgrav}
\eea
With the currents and the propagators we can calculate the amplitude 
which is given by the following relation 
\be
{\cal A}= \frac{6k_{0}^{2}}{\k_{\bot}^{2}}T^{2}_{p}V_{p+1}.
\label{amplitfieldtheory}
\ee
If we compare the amplitude obtained from string calculations
(\ref{closedstringfinampl}) with the amplitude computed from field theory
(\ref{amplitfieldtheory}) we obtain
\be
V_{p+1}\frac{24 \pi}{2^{10}}(4 \pi^{2} \a^{'} )^{11-p} G_{25-p}(Y)= 
\frac{6 \k_{0}^{2}}{\k_{\bot}^{2}}T^{2}_{p}V_{p+1}.
\label{comparedamplit}
\ee
The value of the Green's function that enter the l.h.s. of the relation
(\ref{comparedamplit}) is found in r.h.s. in the momentum space
\be
 G_{25-p}(Y)=\frac{1}{\k_{\bot}^{2}}.
\label{greensinmom}
\ee
The rest of the terms in (\ref{comparedamplit}) give us an equation from 
which we can determine the tension of the brane
\be
\frac{ \pi}{2^{8}\k_{0}^{2}}(4 \pi^{2} \a^{'} )^{11-p} =T^{2}_{p}.
\label{tension}
\ee
We note that there exist a relation between the tension of different branes
given by the following relation
\be
\le (\frac {T_{p+1}}{T_{p}} \rh )^{2}= 4 \pi^{2} \a^{'}.
\label{relatens}
\ee

\subsection{Exercises}

{\bf Exercise 1}\\
Compute the integral over $p$'s in (\ref{amplitopenstring2}).

{\bf Exercise 2 }\\
Compute the trace in (\ref{traceosc}).

{\bf Exercise 3 }\\
Construct the corresponding trace for fermionic oscillators and
compute it.

{\bf Exercise 4 }\\
Show the r.h.s. of (\ref{leadingint}).

{\bf Exercise 5 }\\
Find the action (\ref{gaugefixedgravdil}).

\vskip 2cm


\section{Boundary state description of bosonic $Dp$-branes}

The tree level diagram in closed string theory describes the following
phenomenon: a closed string is generated from the vacuum, it propagates a 
certain interval of time and then it is annihilated again in the vacuum.
One can sandwich this diagram between two states
which will be inserted in the position of the ending circles of the cylinder,
i. e. on the boundary of the world-sheet. Such of states that
describe the creation and annihilation of the closed strings are called
{\bf boundary states}. In the previous paragraph we encountered cylinder
diagrams which described the interaction between two $Dp$-branes at tree 
level. It is then natural to ask if there is any boundary state that could
be interpreted as a $Dp$-brane? The answer is yes. Such of boundary state
represents a microscopic description of the brane in terms of closed string 
modes \cite{pdv1,pdv2}. 

We recall that the open string boundary conditions that define a $Dp$-brane
are given by the following relations
\bea
\6_{\s}X^{a}|_{\s=0} & = & 0,~~~a = 0,1,\ldots ,p 0\nonumber\\
X^{i}|_{\s =0} & = & y^{i}, ~~~i=p+1, \ldots ,25 
\label{bcopenstr}
\eea
To pass to the closed string boundary condition, one has to interpret the 
cylinder as tree-level diagram in closed string sector like in the previous 
section. The relations (\ref{bcopenstr}) take the following form
\bea
\6_{\t}X^{a}|_{\t=0}&=&0,~~~a=0,1,\ldots ,p\nonumber\\
X^{i}|_{\t =0}&=&y^{i},~~~i=p+1,\ldots\ 25.
\label{bcclosedstr}
\eea

If we want to interpret the $Dp$-branes as boundary states, then we must 
implement the boundary conditions (\ref{bcclosedstr}) in the Fock space of
perturbative closed string. This is done by interpreting the string
coordinates as operators 
\bea
\6_{\t}X^{a}|_{\t=0}|B>&=&0,~~~a=0,1,\ldots ,p\nonumber\\
(X^{i}|_{\t =0}- y^i)|B>&=&0,~~~i=p+1,\ldots\ 25.
\label{bchilbert}
\eea
The equation (\ref{bchilbert}) define the boundary state $|B>$. To find its
solution we expand the string operators in terms of oscillation modes using
the solution of the equations of motion given in Section 2
\be
X^{\m}(\t ,\s )=x^{\m} + 2 \a^{'} p^{\m} \t + i \sqrt {\fr{\a^{'}}{2}} 
\sum^{\infty}_{n \neq 0} \le [\a^{\m}_{n}e^{-2in(\t - \s)}+
\tilde{\a }^{\m}_{n}e^{-2in(\t + \s)} \rh ],
\label{oscexpansion}
\ee
which act on the closed string vacuum
\be
|0>=|0>_{\a}|0>_{\tilde{\a}}|p>.
\ee
The equations (\ref{bchilbert}) take the following form
\bea
(\a^{a}_{n} + \tilde{\a}^{a}_{-n})|B>&=&0    \nonumber \\
(\a^{i}_{n} - \tilde{\a}^{i}_{-n})|B>&=&0    \nonumber \\
\hat{p}^{a}|B>&=&0   \nonumber \\
(\hat{x}^{i} - y^{i})|B>&=&0.
\label{bcosc}
\eea
It is worthwhile to note that (\ref{bcosc}) are not the only conditions that
should be imposed on the Hilbert space. Actually, we have to produce 
{\em physical} boundary states, and therefore the negative norm state should
be excluded from the solutions of (\ref{bcosc}). This can be achieved by
taking into account the BRST invariance of the theory which is encoded in the
right and left-moving BRST operators $Q$ and $\tilde{Q}$, respectively. The
BRST invariant state must satisfy 
\be
(Q + \tilde{Q})|B>=0
\label{brstcond}
\ee
which allows us to factorize the boundary state in 
\be
|B>=|B_{X}>|B_{gh}>.
\label{brstfact}
\ee
The ghost contribution to the boundary states is defined in terms of the modes
of ghost and antighost fields by the following
equations 
\bea
(c_{n}+ {\tilde{c}}_{-n})|B_{gh}>&=&0\nonumber\\
(b_{n}- {\tilde{b}}_{-n})|B_{gh}>&=&0.
\label{ghostcontrib}
\eea
We are not going to enter into the details of the BRST quantization and we 
refer the reader to \cite{pdv1}. 

\vskip 0.5cm
{\bf A simpler system}
\vskip 0.5cm

In order to find the solutions to (\ref{bcosc}) we take a look at a simpler
model, a single oscillator and its copy with the following boundary conditions 
\bea
(a \pm \ti{a}^{\dag})|b>&=&0 \non
(a^{\dag} \pm a)|b>&=&0.
\label{singleosc}
\eea
We know that the coherent states of the single oscillator satisfy the
following relations
\bea
a|\a>&=&\a |\a> \non
|\a>&=&e^{\a a^{\dag}}|0>.
\label{coherosc}
\eea
The first thing we can try is to replace the phase $\a$ by an operator that 
depends on the operator $\tilde{a}^{\dagger}$. The simplest phase is this
operator itself multiplied by an unknown phase number, and thus the boundary
state $|b>$ can be written as 
\be
|b>=e^{fa^{\dag} \tilde{a}^{\dagger}}|0>.
\label{boundoscsol}
\ee
The phase $f$ cand be determined by plugging (\ref{boundoscsol}) into 
(\ref{coherosc}) from which we get $f= \pm 1$.

\vskip 0.5cm
{\bf Bosonic $Dp$-brane solution}
\vskip 0.5cm

It is now straightforward to compute the solution to (\ref{bcosc}). To this
end we recall that the creation operators of string modes are given by
\be 
a^{\m \dag}_{n}=\sqrt{n} \a^{\m}_{-n},~~~ n > 0
\label{opstroscdag}
\ee
and that the string is actually a collection of oscillators. The last 
two equations in (\ref{bcosc}) will just localize the state in the transverse
space and thus will give a delta-function factor. The boundary states has
the following general form
\be
|B_{X}>=N_{p}\d^{25-p}(\hat{x}^{i}-y^{i}) \le ( \prod^{\infty}_{n=1} 
e^{- \fr{1}{n} \a_{-n} \cdot S \cdot \ti{\a}_{-n}} \rh )|0>_{\a} |0>_{\tilde{a}}|p=0>,
\label{Dpboundstate}
\ee
where $N_p$ is a normalization constant that should be determined. 
$S$ has the following form
\be
S = (\eta^{ab},-\delta^{ij}).
\label{smatrix}
\ee 
For completeness we write down the ghost contribution (\cite{pdv1})
\be
|B_{gh}>=\exp{\le [\sum^{\infty}_{n=1}( c_{-n} \ti{b}_{-n} - 
b_{-n} \ti{c}_{-n})\rh ]} \le (\fr{c_{0}+ \ti{c}_{0}}{2} \rh ) 
|q=1> |\tilde{q}=1>
\label{Dpghost}
\ee
where the ghost ground state is defined by the following equations
\bea
c_{n}|q=1>&=&0,~~~ n \geq 1 \nonumber\\
b_n|q=1>&=&0,~~~ m \geq 0.
\label{ghostground}
\eea
In any physical gauge we do not have to worry about the unphysical degrees of freedom 
of string theory. An example of such of gauge is the light-cone gauge where one
deals only with the physical degrees of freedom of strings at the cost of
loosing the Lorentz invariance. In the case of eq. (\ref{Dpboundstate}) the light-cone
gauge implies summation over the 24 transverse directions on which the metric is 
Euclidean. 

\vskip 0.5cm
{\bf Computation of $N_p$}
\vskip 0.5cm

In order to have a complete knowledge of the $Dp$-brane state we have to 
compute the normalization constant $N_p$. This can be done by comparing the
interaction amplitude computed in the closed string channel with the result
obtained from the open string channel. In the previous section we obtained the
following result in the open string channel
\be
{\cal A}_{open} = V_{p+1} (8 \pi^{2}\a^{'})^{- \frac{p+1}{2}} 
\int ^{\infty}_{0} \frac{d \t}{\t} \t^{\frac{-p+1}{2}} 
e^{-\frac {Y^{2} \t}{2 \pi \a^{'}}} \left [f_{1}(e^{-\pi \t}) \right ]^{-24}.
\label{openstringchannel}
\ee
In the closed string channel we are at one-loop level. When computing 
the amplitude we have to care only about the propagator of closed strings
between two boundary states
\be
{\cal A}_{closed} =-\fr{1}{2} \mbox{Tr} \log{Z_{0}}=-\fr{1}{2} \mbox{Tr} \log{D}=
<B_{X}|D|B_{X}>.
\label{computeamplitclosed}
\ee
The closed string propagator can be written in complex coordinate on the
Euclidean world-sheet as \cite{gsw}
\be
D=\fr{\a{'}}{4 \pi} \int_{|z| \leq} \fr{d^{2}z}{|z|^{2}}z^{L_{0}-1} 
\bar{z}^{\ti{L}_{0}-1}.
\label{closedpropag}
\ee
where $L_{0}$ and $\ti{L}_{0}$ are the zero mode Virasoro operators for 
right/left modes of closed string given by
\bea
L_{0}&=&\fr{\a^{'}}{4} \hat{p}^{2} + \sum^{\infty}_{n=1} \a_{-n} 
\cdot \a_{n}\nonumber\\ 
\ti{L}_{0}&=&\fr{\a^{'}}{4} \hat{p}^{2} + 
\sum^{\infty}_{n=1} \ti{\a}_{-n} \cdot \ti{\a}_{n}.
\label{lsclosed}
\eea
The propagator (\ref{closedpropag}) has the property that it propagates states
that satisfy the mass-shell conditions $(L_0 -1) |\phi > = 
({\ti{L}}_0 -1)|\phi> =0 $.

The amplitude (\ref{openstringchannel}) factorizes in a trace over zero modes
and a trace over oscillators. Similar computations were performed in the 
previous section and we have found the following values of the two factors
\bea
A_{0}&=&V_{p+1} (2 \pi^{2}\a^{'}t)^{\fr {25-p}{2}}
e^{\fr{-Y^{2}}{2 \pi \a^{'}t}} \nonumber\\
A_{1}&=&\prod^{\infty}_{n=1} \le (\fr {1}{1-|z|^{2n}} \rh )^{24},
\label{traces}
\eea
where the contribution of the ghosts has already been taken into account. Here,
the following notations have been used
\be
|z|=e^{-\pi t},~~~ dzd\bar{z}=-\pi e^{-2\pi t}dtd\s
\label{zetz}
\ee
Now we put everything together and obtain the interaction amplitude in the
closed string channel
\be
{\cal A}_{closed} =N^{2}_{p} V_{p+1} \fr{\a^{'} \pi}{2} (2 \pi \a^{'})^{- 
\frac{25-p}{2}} \int ^{\infty}_{0} \frac{d \t}{\t} \t^{12-\frac{p+1}{2}} 
e^{-\frac {Y^{2} \t}{2 \pi \a^{'}}} \left [f_{1}(e^{\fr{-\pi}{ \t}}) 
\right ]^{-24}.
\label{closedstringA}
\ee
This result should be compared with (\ref{openstringchannel}). To this
end we perform a modular transformation $t \rightarrow 1/\t$ 
\be
{\cal A}_{open} = V_{p+1} (8 \pi^{2} \a^{'})^{- \frac{p+1}{2}} 
\int ^{\infty}_{0} \frac{d \t}{\t} \t^{12-\frac{p+1}{2}} 
e^{-\frac {Y^{2} \t}{2 \pi \a^{'}}} 
\left [f_{1}(e^{-\frac{\pi}{ \t}}) \right ]^{-24}.
\label{closedstringchannel}
\ee
Finally, by comparing (\ref{openstringchannel}) with 
(\ref{closedstringchannel}) we obtain the following value of the normalization
constant
\be
N_{p}=\fr{T_{P}}{2},
\label{normconst}
\ee
where $T_p$ is the brane tension obtained in the previous section.

For further details concerning the boundary state approach to the $D$p-brane
we refer to the pedagogical lecture notes \cite{pdv1,pdv2,mg}. In the original
papers \cite{flprs,vflprs} the relation between the normalization of the 
boundary states and the tension of $D$-branes was established while in 
\cite{bvflprs} the $D$-brane states were constructed in the RNS formalism.
This approach gives a 
good control on the $D$-branes in the limit where it applies. It is useful for
microscopic descriptions of branes as is the case of an alternative formulation
of $D$-branes at finite temperature in the framework of thermo field theory 
proposed in \cite{ivv1,ivv2,ivv3}.

\subsection{Exercises}

{\bf Exercise 1}\\
Construct the propagator (\ref{closedpropag}). Argue its form.

{\bf Exercise 2}\\
Using (\ref{lsclosed}) in (\ref{closedpropag}) obtain (\ref{traces}).

\vskip 2cm


\section{$D$p-branes in Type II theories}

In this section we are going to review basic topics on
supersymmetric $D$-branes. Some of these ideas will be used in 
the following lectures at this school. However, due to the lack of time and
space and because of the complexity of the topics, we are giong to be rather
quick. We refeer the interested sudents to the very good reviews 
\cite{cj,jp,pdv1,pdv2,superD}.

\subsection{Closed RNS Superstring}

The bosonic string theory presented above suffers from some serious drawbacks
as the presence of tachyons in the perturbative spectrum and the absence of
fermions. One way to introduce the fermions in string theory is through 
{\em supersymmetry} which is a symmetry of the original classical theory that
transforms the bosons into fermions and vice-versa. 
From a
pragmatic point a view, it is known from field theory that supersymetry can 
cure the divergencies. The supersymmetry can be constructed either by 
supersymmetrizing the world-sheet fields (Ramond-Neveu-Schwarz) or
by constructing the target-space action (Green-Schwarz). The two constructions
are equivalent in the light-cone gauge. The equivalence is based on the modular
symmetry and is implemented by the Gliozzi-Scherk-Olive projection which
projects out of spectrum half of the states of RNS string. What is left is
a theory with space-time supersymmetry equivalent to GS string. Through the GSO
projection the tachyon is killed so that the vacua of superstrings are stable
\cite{gsw,jp}.

\vskip 0.5cm
{\bf Classical closed superstring}
\vskip 0.5cm

The action of the superstring in the RNS formulation is 
\be
S = \f{1}{4\pi\a '}\int_{\Sigma} d^2 \s (\6_{\a}X^{\m}\6^{\a}X_{\m}-
i\bar{\psi}^{\m}\r^{\a} \6_{\a}\psi_{\m}),
\label{actRNS}
\ee
where to each bosonic field $X^{\m}(\s)$ it is associated a two-dimensional
Majorana fermionic field $\psi^{\m}_A(\s)$ with $A=1,2$. The Dirac matrices
in two dimensions can be chosen imaginary \cite{gsw} and by definition they 
should satisfy the following algebra
\be
\{ \r^{\a}, \r^{\b} \} = - 2 \eta^{\a \b}.
\label{diracalg}
\ee
The action (\ref{actRNS}) has the Poincar\'{e} symmetry in target-space, 
super-Weyl invariance in two-dimensions and supersymmetry. The supersymmetry
transformations mix the bosonic and fermionic variables and are given by
the following relations
\bea
\delta X^{\m} &=& \bar{\e}\psi^{\m}\nonumber\\
\delta \psi^{\m} &=& -i\r^{\a}\6_{\a}X^{\m}\e ,
\label{susy2d}
\eea
where $\e$ is a infinitesimal constant Majorana spinor in two-dimensions that 
parametrizes the supersymmetry transformations. The following supercurrent
corresponds to the supersymmetry
\be
J_{\a} = \f{1}{2}\r^{\b}\r_{\a}\psi^{\m}\6_{\b}X_{\m}.
\label{supercurrent}
\ee
As in the pure bosonic case, the system is subject to constraints. The constraints are 
the equations of motion for the two-dimensional supergravity fields 
(graviton and gravitino) if a 
general world-sheet metric is considered. However, in the superconformal gauge
in which the action (\ref{actRNS}) is written, they should be imposed by hand 
and they have the following form
\bea
T_{\a\b} &=& \6_{\a}X^{\m}\6_{\b}X_{\m} + \f{i}{2}\bar{\psi}^{\m}
\r_{(\a}\6_{\b)}\psi_{\m}
-\f{1}{2}\eta_{\a\b}(\6_{\g}X^{\m}\6^{\g}X_{\m} + \f{i}{2}\bar{\psi}^{\m}
\r^{\a}\6_{\a}\psi_{\m}) = 0\\
J_{\a}  &=&  0.
\label{superconstraints}
\eea   
Also, the super-Weyl transformations imply that
\bea
T^{\a}_{\a} &=&0\\
\r^{\a}J_{\a} &=&0.
\label{superweylconstr}
\eea
It is easy to show that the equations of motion from (\ref{actRNS}) are the
two-dimensional wave equation and the Dirac equation, respectively
\bea
\6^{\a}\6_{\a}X^{\m} &=&0\\
\r^{\a}\6_{\a}\psi^{\m}&=&0.
\label{supereqsmotion}
\eea
The topology of the string determines the boundary conditions that should be
imposed on the bosonic coordinates. For fermionic coordinates, the boundary
conditions are determined from both topology of world-sheet and the 
supersymmetry. For bosonic coordinates, the boundary conditions are the same
as in eq.(\ref{boundclosed}) for closed string and eq.(\ref{boundopen}) for
open string, respectively. For fermions, there are two boundary conditions that
can be imposed either after the fermions performs a complete period on the
closed string or at the two ends of the string for open strings.
These boundary conditions simply state that the spin of the fermion can flip.
Since we focus on the closed strings, the boundary conditions are given
by the following relations
\bea
\psi^{\m}(\t,\s + \p ) &=& + \psi^{\m}(\t,\s )~~~R~~b.c.\\
\psi^{\m}(\t\s+\p ) &=& - \psi^{\m}(\t,\s )~~NS~b.c.
\label{fermionicbc}
\eea 
One can use the powerful techniques of complex analysis and of conformal fied 
theories in two dimensions to study the
superstring 
if we  map the cylinder into the 
comples plane $\cal{C}^*$
\be
z= e^{2(\t -i\s )} = e^w,
\label{complexplane}
\ee
from which we can see that the boundary conditions can be written in the 
following form
\bea
\psi^{\m}(e^{2\p i}z) &=& - \psi^{\m}(z)~~~R~~b.c.\\
\psi^{\m}(e^{2\p i}z) &=& + \psi^{\m}(z)~~NS~b.c..
\label{complexbc}
\eea
To prove the relations above, one uses the periodicity on the cylinder of
the holomorphic (and antiholomorphic) fermions.

\vskip 0.5cm
{\bf Massless spectrum of the closed superstring}
\vskip 0.5cm

In order to identify the excitations of the massless fields, one has to 
quantize the superstring. One can apply exactly the same methods used for
the bosonic string \cite{gsw,jp}. We are going to use the canonical 
quantization
for pedagogical resons.

The bosonic coordinates $X^{\m}(\s)$ were quantized in Section 2. To quantize 
the fermionic coordinates we note that the left- and right-moving modes on
the closed string are independent. Consequently, in the complex plane, the
solutions to the Dirac equation decompose into holomorphic and anti-holomorpic
parts. The Fourier decomposition of the holomorphic part is 
given by the following relations
\bea
\psi^{\m}(z)&=& \sum_{n\in \cal{Z}} d^{\m}_{n} z^{-n-1/2}~,~~~(R)\\
\psi^{\m}(z)&=& \sum_{r\in Z'} b^{\m}_{r} z^{-r-1/2}~,
~~~(NS)
\label{holompsi}
\eea
where $Z' = Z + 1/2$.
Similar expressions hold for anti-holomorphic part. The holomorphic and
anti-holomorphic sectors describe the left- and right-moving modes, 
respectively. The coefficients of the expansion in (\ref{holompsi}) are 
interpreted as operators acting on the Fock space which is given by the 
tensor product of left-and right modes
\be
| \mbox{left} > \otimes | \mbox{right} >.
\label{tensorFock}
\ee
The interpretation of coefficients in terms of creation and annihilation
operators is given by their algebra which can be obtained from the 
postulated anti-commutator relations among fermionc fields \cite{gsw,jp}
\bea
\{ d^{\m}_{m}, d^{\n}_{n} \} &=& 
\eta^{\m\n}\d_{m+n,0}~,~~~(R)\\
\{ b^{\m}_{r}, b^{\n}_{r} \} &=& \eta^{\m\n}\d_{r+s,0}~,~~~(NS)
\label{anticomm}
\eea 
and similarly for right-moving modes. From the relations above we see that
one can have $\d_{m+n,0}=1$ in the Ramond sector.
Thus, the states $d^{\m}_{0} |0>$ are in the massless representation
of the Clifford algebra
\be
\{ d^{\m}_{0}, d^{\n}_{0} \} = \eta^{\m \n}.
\label{clifford}
\ee
This means that {\em the ground state in the R-sector is a spinor}. 

The super-Virasoro of the theory is lost through quantization but one can
show that the anomaly that appears can be cancelled if the dimension of the
space-time is $D=10$. In ten-dimensions, the background spinor from the 
R-sector can have both $\pm$ chiralities and it is a Majorana-Weyl spinor
which we denote by $ |A^+>$ and $|A^->$, respectively, where $A$ are spinor
indices of $Spin(8)$ transversal group.

The above analysis can be repeated verbatim for the right-moving sector.
Due to the tensor product structure of the Fock space (\ref{tensorFock})
we can have background spinors of different chiralities in the two sectors.
Therefore, we can classify the closed superstrings in 
\be
\begin{array}{ccc}
\mbox{Theory}&~~~~ &\mbox{Ground state}\\
\mbox{Type IIA} &~~~~ &| A^+ > \otimes |\bar{ A^-} >\\
\mbox{Type IIB} &~~~~ &| A^+ > \otimes |\bar{ A^+} > 
\end{array}
\label{theories}
\ee
The Type IIA is not chiral, i.e. the vacua in the two sectors have opposite
chirality while the Type IIB theory is chiral since the two vacua has the same
chirality. (The sign between the two spinors is relative.)

We can quantize the system along the same line as in the bosonic case, but
we are not going to present the details here. They can be found in the 
textbooks \cite{gsw,jp}. The mass operators in the holomorphic sector, 
necessary to classify the spectrum
of the superstring, are given by the following relations 
\bea
\f{1}{4}M^2 &=& \sum_{n>0} \a^{i}_{-n}\a^i_n + 
\sum_{r>0}rb^{i}_{-r}b^{i}_{r} -\f{1}{2}~,~~~(R)\\
\f{1}{4}M^2 &=& \sum_{n>0}\a^{i}_{-n}\a^{i}_{n} + 
\sum_{n>0}nd^{i}_{-n}d^{i}_{n}~,~~~~~(NS)
\label{masslightcone}
\eea
where $\a '=1$ and the indice $i=1,2,\ldots,8$ labels the transverse 
directions (light-cone gauge.) The 
one-half factor from the $NS$-sector comes from the normal ordering of the
super-Virasoro operator. In the $R$-sector its value is zero. 

The states are constructed by acting with the creation operators on the vacuum.
We are interested in space-time supersymmetric states which is the observed 
supersymmetry, rather than in the world-sheet supersymmetric states. To obtain
the target-space spectrum, we have to perform the GSO projection onto the
world-sheet vector space. This projection is represented by the GSO operator
$(-)^F$ under which the bosonic fields $X^{\m}$ are even and the fermionic
ones $\psi^{\m}$ are odd
\be
[ (-)^F, X^{\m}] = 0~,~~~~~\{ (-)^F, \psi^{\m} \} =0.
\label{GSOop}
\ee  
These properties determine the operator $F$ up to a sign which is fixed by 
asking that in the open superstring spectrum the photon be invariant under
the GSO projection. 

The GSO operator is represented by
\bea
\Gamma &=& \G^0 \cdots \G^8 (-)^{\sum_{n>0}d^{i}_{-n}d^{i}_{n}}~,~~~(R)
\\
G &=& (-)^{F+1}= (-)^{\sum_{r>0}b^{i}_{-r}b^{i}_{r} + 1}~,~~~(NS) 
\label{GSOoperators}
\eea
in the two sectors of holomorphic spectrum. 
Here, $\G^i$ are the Dirac matrices in $D=10$ dimensions.
With these conventions, the states
that are odd under the action of the GSO operators are projected out. 

In the light cone gauge the massless spectrum of the closed superstrings is 
given by the product between the left- and right-moving states classified
according to the three representations of the $SO(8)$ group (the little group
of $SO(1,9)$), namely $\mbox{\bf 8}_{\mbox{\bf v}}$, 
$\mbox{\bf 8}_{+}$,
$\mbox{\bf 8}_{-}$ as follows
\be
(\mbox{\bf 8}_{\mbox{\bf v}} \oplus \mbox{\bf 8}_{+})_{\mbox{l}}
\otimes
(\mbox{\bf 8}_{\mbox{\bf v}} \oplus 
\mbox{\bf 8}_{\pm})_{\mbox{r}},
\label{specrepropt}
\ee
where the subscripts {\bf l} and {\bf r} stand for left- and right-moving
modes, respectively. 

\vskip 0.2cm
{\bf Type IIA massless spectrum}
\vskip 0.2cm

The non-chiral closed superstring has the following spectrum
\bea
(\mbox{\bf 8}_{\mbox{\bf v}} \oplus \mbox{\bf 8}_{+})_{\mbox{l}}
\otimes
(\mbox{\bf 8}_{\mbox{\bf v}} \oplus 
\mbox{\bf 8}_{-})_{\mbox{r}}&=&
(\mbox{\bf 1} + \mbox{\bf 28} + \mbox{\bf 35}_{\bf v})_{NS-NS}
\oplus  (\mbox{\bf 8}_{\mbox{\bf v}} + 
\mbox{\bf 56}_{\mbox{\bf v}})_{R-R}\nonumber\\
&& \oplus 
(\mbox{\bf 8}_{+} + 
\mbox{\bf 56}_{-})_{NS-R}
\oplus
(\mbox{\bf 8}_{-} + 
\mbox{\bf 56}_{+})_{R-NS},
\label{specIIA}
\eea
The states from $NS-NS$ and $R-R$ sectors are bosonic since they are either 
a product
of two bosonic states, or a product of two spinor states while the states 
from the $NS-R$ and $R-NS$ sectors are fermionic being products of a bosonic
and a fermionic state. The number of bosonic and fermionic degrees of freedom
match and this is the first indication of the existence of supersymmetry
\cite{gsw}. The numbers in the brackets indicate the irreducible
representations of the $SO(8)$ group. According to it, the bosonic states
can be identified with the excitations of the dilaton, Kalb-Ramond field and
gravitational potential $\phi, B_{\m\n}, g_{\m\n}$ in the $NS-NS$ sector and
with the excitations of an one-form and a three-form fields $A_{\m}$ and
$A_{\m\n\r}$, respectively, in the $R-R$ sectors.

\vskip 0.2cm
{\bf Type IIB massless spectrum}
\vskip 0.2cm

The chiral closed superstring has the following spectrum
\bea
(\mbox{\bf 8}_{\mbox{\bf v}} \oplus \mbox{\bf 8}_{+})_{\mbox{l}}
\otimes
(\mbox{\bf 8}_{\mbox{\bf v}} \oplus 
\mbox{\bf 8}_{-})_{\mbox{r}}&=&
(\mbox{\bf 1} + \mbox{\bf 28} + \mbox{\bf 35}_{\bf v})_{NS-NS}
\oplus 
(\mbox{\bf 1} + \mbox{\bf 28} + \mbox{\bf 35}_{+})_{R-R}
\nonumber\\
&& \oplus 
(\mbox{\bf 8}_{-} + 
\mbox{\bf 56}_{+})_{NS-R}
\oplus
(\mbox{\bf 8}_{-} + 
\mbox{\bf 56}_{+})_{R-NS},
\label{specIIB}
\eea
The bosonic fields that correspond to the massless irreducible representations
in the Type IIB theory are the dilaton, Kalb-Ramond field and
gravitational potential $\phi, B_{\m\n}, g_{\m\n}$ in the $NS-NS$ sector
and a scalar field, a two-form field and a self-dual four-form field 
$\chi$, $A_{\m\n}$ and $A^{+}_{\m\n\r\s}$ in the $R-R$ sector. The self-duality
of the four form field means that the field strength is equal to its Hodge 
dual.

As we can see from (\ref{specIIA}) and (\ref{specIIB}) above, the graviton 
appears in the two theories. Also, by GSO-projection the tachyon has been removed
since it is and odd state under the action of $(-)^F$ operator. The theories
are free of anomalies in $D=10$, free of tachyons and contain fermions. Beside
the fundamental interactions, some other interactions mediated by $p$-form 
fields are predicted. In four dimensional space-time such of interactions 
cannot be written since there is not enough room to accomodate the higher
rank $p$-forms. Indeed, a three-form in four dimensions has the same number of
components as a two-form, and a four-form as a one-form or a vector. Thus, no
new gauge potentials can be constructed. The Type II theories have $N=2$
supersymmetry in $D=10$, corresponding to the two generators of opposite
or equal chirality.  

Let us note in the end that there are three more string theories known as 
Type I,
Heterotic $SO(32)$ and Heterotic $E_8 \times E_8$. The Type I theory
contains opens strings. Therefore, it has just $N=1$ supersymmetry. The 
heterotic theories are supersymmetric just in one of the sectors 
(left or right). The dilaton, Kalb-Ramond and gravitational fields are common
to all these theories, but the $p$-forms differ from one theory to another.
For more details, we refer the reader to \cite{gsw,jp,cj}.

\vskip 0.5cm
{\bf Spin operators and space-time supercharges}
\vskip 0.5cm

All the nice features of the superstrings come from the new symmetry, the
supersymmetry that has been introduced in (\ref{susy2d}). The supersymmetry is
implemented at the quantum level by the supercharge operators. 

Let us review
how they can be constructed \cite{gsw,jp}.
In the complex world-sheet variables, the superstring theories are described
by {\bf superconformal field theories} (SCFT). These field theories are well 
known.
They have nicer properties than field theories in four dimensions since there
are an infinite number of symmetries in two-dimensions (Virasoro) that give
a good control of the $S$-matrix. Using the techniques of SCFT, one can
show that the spinor ground states in the $R$-sector can be obtained by acting
with some operators called {\bf spin operators} $S^{\pm}_a(z)$ and 
$\bar{S}^{\pm }_a (\bar{z})$ on the vacuum of the $NS$-sector in 
both left- and 
right-moving sectors \cite{jp,cj}. 
The spin operators transform as 
space-time spinors which justify their names and there are 32 of them.
One fundamental object in conformal field theory is the 
{\bf operator product expansion} (OPE) which encodes all the information 
about the theory since it is equivalent with the comutation relations. For
the $S$ and $\bar{S}$ fields, the basic OPE's are with the fermionc fields
\bea
\psi^{\m}(z)S(w) &\sim & (z-w)^{-\f{1}{2}}\G^{\m}S(w)\\
\bar{\psi}^{\m}(\bar{z})\bar{S}(\bar{w}) 
&\sim & (\bar{z}-\bar{w})^{-\f{1}{2}}\G^{\m}\bar{S}(\bar{w}), 
\label{OPE}
\eea
where $\G^{\m}$ are the Dirac matrices and $\sim$ means that the irregular 
terms are discarded.
The contour integral of fermion-emission operators are just the supercharges
\be
Q = \oint \f{dz}{z}S(z)~,~~~~\bar{Q} = -\oint \f{d\bar{z}}{\bar{z}}
\bar{S}(\bar{z}).
\label{supercharges}
\ee 
This way of understanding the supercharges will be useful later when we will 
analyse the supersymmetries preserved by the $D$-branes. 

\subsection{Type II supersymmetric $D$-branes }

As in the case of the bosonic string theory, the $D$-branes in the superstring
theory are defined by a mixed Neumann and Dirichlet boundary conditions on the
open superstring world-sheet. The fermionic boundary conditions should be 
compatible with the supersymmetry
(\ref{susy2d}). If we pass to complex coordinates on the Euclidean complex 
plane $z = \exp(\t + i\s)$ the whole set of boundary conditions can be written
as follows \cite{jp,superD}
\bea
\6 X^{a} &=& \bar{\6} X^{a}|_{Im \; z = 0}\nonumber\\
\6X^{i} & = & - \bar{\6} X^{i}|_{Im \; z=0}
\label{bosonicsuperbc}
\eea
for bosonic coordinates, and
\bea
\psi^{a} &=& \bar{\psi}^a|_{Im\;z=0}~,~~\psi^{i} =- \bar{\psi}^i|_{Im\;z=0}~,
~~(R)\\
\psi^{a} &=& -\bar{\psi}^a|_{Im\;z=0}~,~~\psi^{i} = \bar{\psi}^i|_{Im\;z=0}~,
~~(NS)
\label{fermionicsuperbc}
\eea
for fermionic coordinates, where $a=0,1,\ldots ,p$ and $i=p+1,\ldots ,9$.

In general, the presence of such of extended objects in superstring theory
will break the original symmetries. We saw that in the bosonic theory where
the Poincar\'{e} symmetry was broken.  In the supersymmetric case
$D$-brane breaks the space-time symmetry down to 
$SO(1,p)\times SO(9-p)$. We may ask what happens with the supersymmetry?

In order to have some conserved supersymmetry we need supercharges that leave 
the vacuum invariant, or equivalently, spin fields. Since they  
satisfy the OPE (\ref{OPE}) it is easy to see that at the boundary
there will be some relations that should be imposed on the spin fields in
order to mantain the compatibility between the boundary conditions 
(\ref{fermionicsuperbc}) and the OPE (\ref{OPE}). These relations represent
the boundary conditions for spin fields and one can show that they transform
$S$ into $\bar{S}$ and vice-versa \cite{jp,superD}
\be
S = \Pi_{(p)}\bar{S}.
\label{spinfieldbc}
\ee
One can look for an operator $\Pi_{(p)}$ that is constructed from Dirac 
matrices since the spin operators transform as spinors in $D=10$. By 
introducing (\ref{spinfieldbc}) into (\ref{OPE}) one can show that $\Pi{(p)}$
should satisfy the following relations \cite{jp,superD}
\be
[ \Pi_{(p)}, \G^{a} ] = 0~,~~~\{ \Pi_{(p)},\G^i \} = 0.
\label{relPp}
\ee
Such of operator exists, and has the following form
\be
\Pi_{(p)} = i^{9-p}\G_{11}\G^{p+1}\G_{11}\G^{p+2}\cdots\G_{11}\G^{9},
\label{Ppoperator}
\ee
where 
\be
\G_{11}=\G^0\G^1\cdots\G^9.
\label{G11}
\ee
Since $\Pi_{(p)}$ maps left spin operators to right spin operators, it should 
flip the chirality for Type IIA spin operators and leave it invariant for 
Type IIB spin operators. This chirality is flipped for $p$ even, and left 
unchanged for $p$ odd. Therefore, we have the following supersymmetric 
(or BPS) $D$-branes
\be
\begin{array}{ccl}
\mbox{Theory}&~~~&D\mbox{p-branes}\\
\mbox{Type IIA}& ~~~&p=0,2,4,6,8\\
\mbox{Type IIB}& ~~~&p=-1,1,3,5,7,9
\end{array}
\label{tableDb}
\ee
The $p=-1$ brane makes sense only in the Euclidean space-time where it is 
interpreted as a soliton. $p=9$ is a degenerate case in which the string
can propagated freely in the bulk of space-time and it is consistent only
in TypeI theory where some auxiliary construction should be done.

We can see in this way that the $Dp$-brane break the supersymmetry of the 
background and only half of it is preserved. Some other configurations of 
BPS-branes can be imagined which break the supersymmmetry to 1/2, 1/4,...
of the original number of supercharges. There are also non-BPS branes
which do not preserve any supersymmetry at all, but discussing these topics
is out of the scope of these lectures (see \cite{jp,cj,pdv1,pdv2,superD}.)  

\subsection{Some properties of the $D$-branes}

The supersymmetric $D$-branes can be treated in the similar fashion as the 
bosonic ones studied in the previous sections. Besides their tension, they are
characterized by other physical quantities as RR charges and supersymmetry. We
are going to review the basic properties of BPS-branes in what follows.

\vskip 0.2cm
{\bf Tension and charge of $D$-brane}
\vskip 0.2cm

The tension of the brane is computed from the exchange of the massless modes
of the closed strings. The difference from the bosonic case lies in the fact 
that there is a two-form field in the RR sector whose excitations should be
taken into accout. The details of the computations are given in\cite{jp} 
(see also \cite{cj,superD}) and the result is
\be
T^{2}_{p} = \f{\pi}{K^{2}} (4\pi^{2}\a ')^{3-p},
\label{tenssusy}
\ee 
where $K$ is the Newton's constant in ten dimensions. The tension of the brane 
equals its RR charge-density $e_p$. The RR $p$-form field couples with the 
brane as
the point-like particle couples with the gauge potential in four dimensions. 
The coupling can be electic-like or magnetic-like, i. e. with the strength-form
field or with its dual. For example,  
the electric-like interaction term is given by the Wess-Zumino action
\be
e_{p+1} \int d^{p+1}\xi \hat{A}^{p+1} 
\label{wesszumino}
\ee
in the simplest situation when the topology is kept simple. Here, the integral
is over the world-volume of the brane and $A^{p+1}$ is the pull-back of the
RR field on the world-volume of the brane. $e_p$ is the $RR charge$ of the
$D$-brane and can be calculated by integrating the dual of the field strength
on a sphere in the transverse space arround the brane 
\be
e_{p+1} = \int_{S^{8-p}} *F_{8-p}.
\label{electric}
\ee
The magnetic charge can be computed similarly as
\be
g_{7-p} = \int_{S^{p+2}} F_{p+2}.
\label{magnetic}
\ee
In Eq.(\ref{electric}) and Eq.(\ref{magnetic}) the rank of the forms have
been written explicitely. We recall that in order to perform the integration,
the dimension of the manifold on which we integrate and the rank of the
integrated form should be equal 
(see also K. Stelle's lecture notes at this school.) The electric and magnetic
charges of the $D$-branes can be quantized following Dirac's prescription
\be
\frac{e_{p+1}g_{7-p}}{4 \pi} = \f{n}{2},
\label{Dirac}
\ee
where $n$ is an integer number.

It is important to note that the parallel BPS-branes do not 
feel any force among them since the contribution of the NSNS and RR sectors is
equal and of opposite sign. However, for branes at angles the situation is 
different and for some values of relative angles and distances tachyons
appear in the system \cite{Dangles}. 

\vskip 0.2cm
{\bf Effective action}
\vskip 0.2cm

The effective action of the BPS D-branes can be calculated by using the same
method as in the bosonic case. The difference comes from the RR field which 
for a hyperplanar static brane has only one component coupling with the brane.
However, this action would describe only the bosonic sector of the theory.
In order to obtain a supersymmetric low energy action, one has to generalize
it to include the space-time supersymmetry. The supersymmetric action was
obtained in \cite{suact}. There is some subtelty involved in this 
generalization, given by the fact that the fermionic space-time variables are
twice in number than necessary. Consequently, one has to impose another
local fermionic space-time symmetry called $k-symmetry$, known from the
supersymmetric generalization of particles and strings \cite{gsw,jp}. This
symmetry will ensure the correct space-time degrees of freedom for the
fermionic coordinates.

The bosonic part of the effective action contains two terms 
called Dirac-Born-Infeld action and Wess-Zumino action which have the 
following forms
\bea
S_{DBI} &=& T_{p}\int d^{p+1}\xi 
e^{-\hat{\phi}}\left[ -\mbox{det}({\hat{G}}_{\a\b} + {\hat{B}}_{\a\b} + 
2\pi\a 'F_{\a\b}) \right] ^{\f{1}{2}}\\
S_{WZ} &=& T_{p} \int d^{p+1}\xi \hat{A} \wedge e^{2\pi \a 'F} \wedge 
\left (\f{\cal{A}(\cal{T})}{\cal{A}(\cal{N})} \right) ^{\f{1}{2}}.
\label{susydbiact}
\eea
Here, the space-time fields are pulled-back on the world-volume of the 
$D$-brane. In the Wess-Zumino action, the exponential should be expanded to 
saturate the dimension of the integral. This is because the fields are 
expressed by differential forms. The higher dimensional terms give zero 
contribution. The Dirac-Born-Infeld term has properties similar to the 
effective action of 
the bosonic $D$-brane. As was discussed in Section 3, it generalizes the 
geometric action, i. e. that is proportional to the volume of brane 
trajectory, to a background with non-vanishing fields. The generalization of
DBI-action to non-abelian potentials $A^{a}(\xi )$ is not understood
yet.

Let us discuss the Wess-Zumino term. It is interpreted as the term that
generalizes the coupling of the brane with the RR $(p+1)$-form fields and
should be calculated by expanding the exponential as discussed above.
The objects $\cal{A}(\cal{N})$ and $\cal{A}(\cal{T})$ are topological
invariants that characterize the tangent bundle over the space-time manifold,
decomposed into the normal bundle $\cal{N}$ to the brane world-volume and
the tangent bundle $\cal{T}$ to it. This invariant is called {\it roof-genus}
or {\it Dirac genus} of the bundle and it is defined in terms of the Chern
classes. For a vector bundle $E$, the definition of the roof-genus is
\be
{\cal{A}}(E) = \prod_{n} \frac{\l_n /2}{\sinh (\l_n /2)},
\label{roofgen}
\ee
where $\l_n = c_1 (L_n )$. Here, $c_1$ is the first Chern class of the 
bundle $L_n$ which is a line bundle. One can expand the roof-genus in terms
of Pontryagin classes
\be
{\cal{A}}(E) = 1 - \f{1}{24}p_1 (E) + \cdots,
\label{roff-pnt}
\ee
where
\be
p_n (E) = (-1)^n c_{2n}(E \otimes_R C)
\label{pnt}
\ee
is the $n$-th Potryagin class of the vector bundle $E$ tensored with the 
complex numbers field $C$. More intuitively, the roof-genus can also be 
expressed in terms of the curvature of the 2-form $R$
\be
{\cal{A}}(E) = 1 + \f{1}{(4\pi )^2} \f{1}{12} {\mbox Tr} R^2 + \cdots
\label{roof-cur}
\ee
which defines the Dirac genus as a sum of invariant polynomials in the 
curvature form \cite{geometry}.

The Wess-Zumino action generalizes the coupling of a point-like charge with
a gauge potential to a higher dimensional object coupled to the $(p+1)$-form.
Since the $D$-brane is an extended object, it may assume various topologies in
a given background. Therefor, the coupling terms should be topological 
invariants to guarantee that the formulation of the theory does not change 
when going from one topology to another. This is how the topological invariants
in the action can be roughly explained. In the case of a point-like particle 
wecannot see all these complications due to the trivial topology of the 
particle.


\subsection{Exercises}

{\bf Exercise 1}\\
Prove (\ref{complexbc}) starting from (\ref{fermionicbc}).

{\bf Exercise 2}\\
Show that, for the superstring in trivial background, the GSO operator 
$(-)^F$ projects out the tachyon from the spectrum.

{\bf Exercise 3}\\
Argue that the electric and magnetic charges of a $D$-brane statisfy the
quantization condition (\ref{Dirac}).

\vskip 2cm


\section{Discussions}

In these notes we have argued that there are extended objects in string theory
called $D$-branes which exhibit, beside a geometric structure, physical 
properties as tension and, for supersymmetric branes, charges. We have derived
the effective action of the bosonic $D$-branes and we have given a 
microscopic description of them. Also, we have briefly mentioned some of the
properties of the supersymmetric BPS-branes.

The material presented in these lectures is in some sense ``classic''. We have
not discussed any of the more advanced and new results, part because of the
extended background material needed to understand these topics which is 
unfamiliar to many students and 
partly because of the lack of time. However, there are some exciting ideas 
which we are going to review briefly now.

{\bf BPS D-brane dualities}

We have seen above that the $D$-branes appear in three of the five string 
theories. In Type IIA and Type IIB, the dimension of the world-volume of the
brane is odd, respectively even. However, by compactifying the world-volume
of, say, a Type IIA $Dp$- brane, on a circle of radius $R$ and taking the 
limit $R \rightarrow 0$, one obtaines a manifold with the dimension $p$. If 
this manifold is identified with the world-volume of a $D(p-1)$-brane we 
have a map from Type IIA branes to Type IIB branes. This is a basic way to 
establish relations among string theories by using branes, relations 
called {\em dualities}. Actually, there are technical details in constructing 
the dualities. There are several types of them and in the last years there
have been done many works in this field \cite{dualities}.

{\bf World-volume action of BPS D-branes}

In Section 3 we derived the low energy limit of the action of bosonic 
$D$-branes and in Section 6 we discussed its generalization to 
supersymmetric branes. As was already mentioned, it is not very clear how
to generalize the action to non-abelian gauge potential. This is an important
line of research, and works have been done recently (for a review see 
\cite{tsey}.) The problems are related to ambiguities in the definition of the
expansion of the determinants of strength-tensor for non-abelian fields. 
Also, it is not known how to construct a polynomial and local action for 
$D$-branes that host a self-dual form field on the world-volume. 

Most of the studies have been devoted to low energy limit of the $D$-branes. 
The higher energy form of the action for them is unknown. 
 
{\bf Non-commutativity}

When the Born-Infeld action is generalized to $N$ parallel $Dp$-branes, the
coordinates on the world-volume of the branes become non-commutative functions.
This motivated many works on non-commutativity of both branes and strings
\cite{seibwittnc}. The gauge field on the world-volume is $U(N)$ 
non-commutative gauge theory and the corresponding low energy action is
a non-commutative Born-Infeld action. The implications of non-commutativity of
space-time physics are currently under intense investigation.

{\bf Non-BPS branes}

In the last two years there has been an incresing interest in the $D$-branes
that do not preserve any supersymmetry of the background (see the following 
pedagogical reviews \cite{alr,ass1,ass2}. In brane-antibrane
pair (for more than one pair and non-parallel branes see \cite{Vancea:2001zu})
there is a tachyonic field that cannot be elliminate through the usual GSO
projection. Its effective potential display a local minimum in which it was
conjectured that the system reaches a stable state. The evolution to this 
state is called decayment. By this process non-BPS branes can be obtained from
brane-anti-brane pairs and vice-versa, but the dimension of the branes at
the beginning and at the end of the process are different. This is a new type
of interaction between branes. Since the tachyons are off-shell states, the
best tool to investigate the decayment is string field theory. (For  
references on tachyon physics in $D$-brane theory see \cite{tach}.)

{\bf Classification of branes}

The study of non-BPS branes led to some unexpected applications of mathematics:
the classification of brane charges was shown to be given by the topological 
K-theory of the fibre bundles \cite{ktheory}. This construction was extended
to M-theory in \cite{duliu1,duliu2}. Also, a tentative to include the massive
branes in the K-theory framework wos done in \cite{eualgk}.
More recently, more general 
$D$-brane solutions suggested the derived categories as the most appropriate
framework for describing the brane charges and decayment \cite{moore}.
Treating $D$-branes within the framework of string field theory suggests more
algebraic structure behind brane physics.

We cannot end this section without mentioning two revolutionary results in 
theoretical physics introduced by $D$-branes: the Maldacena's conjecture
and the sub-millimeter extra dimensions. The first one represents a first 
explicit
proposal and tool of mapping between field theory and gravity.
The second theory proposes a solution to the hierarchy problem based on
the idea that the Standard Model is localized on a three brane while the 
gravity lives in the bulk of a five dimensional space-time (see, for
example \cite{mald,pvmal1,pvmal2} and \cite{rs,smc}.)

There are many more things that could be said about $D$-brane theory and many 
other interesting topics that have been investigated recently. The physics
of $D$-branes is far from being understood, but it is clear that $D$-branes
have been helping us to reveal some of the structure of the most interesting 
models
of the high energy physics and it is likely that their role will not be
less important in future.


\begin{thebibliography}{99}

\bibitem{gsw}M.~B.~Green, J.~Schwarz, E.~Witten, {\it Superstring Theory},
(Cambridge University Press, 1989)
\bibitem{jp}J.~Polchinski, {\it String Theory}, (Cambridge University Press,
1999)
\bibitem{nb} N. ~Berkovits, lectures at this school
\bibitem{cj}
C.~V.~Johnson,
hep-th/0007170.
\bibitem{Kiritsis:1998hj}
E.~Kiritsis,
hep-th/9709062.
\bibitem{abou}
A.~Abouelsaood, C.~G.~Callan, C.~R.~Nappi and S.~A.~Yost,
Nucl.\ Phys.\ B {\bf 280}, 599 (1987).
\bibitem{jpprl}
J.~Polchinski,
Phys.\ Rev.\ Lett.\  {\bf 75}, 4724 (1995)
[hep-th/9510017].
\bibitem{ag}
L.~Alvarez-Gaume and M.~A.~Vazquez-Mozo,
hep-th/9212006.
\bibitem{at}A.~A.~Tseytlin,
A.~A.~Tseytlin,
J.\ Math.\ Phys.\  {\bf 42}, 2854 (2001)
[hep-th/0011033].
\bibitem{callan}
C.~G.~Callan, E.~J.~Martinec, M.~J.~Perry and D.~Friedan,
Nucl.\ Phys.\ B {\bf 262}, 593 (1985). SI ALTII TOTI! Chiara Nappi, etc.
\bibitem{call}
C.~G.~Callan, C.~Lovelace, C.~R.~Nappi and S.~A.~Yost,
Nucl.\ Phys.\ B {\bf 288}, 525 (1987).
\cite{Callan:1988st}
\bibitem{Callan:1988st}
C.~G.~Callan, C.~Lovelace, C.~R.~Nappi and S.~A.~Yost,
Phys.\ Lett.\ B {\bf 206}, 41 (1988).
\bibitem{Callan:1988wz}
C.~G.~Callan, C.~Lovelace, C.~R.~Nappi and S.~A.~Yost,
Nucl.\ Phys.\ B {\bf 308}, 221 (1988).
\bibitem{Alvarez-Gaume:1981hn}
L.~Alvarez-Gaume, D.~Z.~Freedman and S.~Mukhi,
Annals Phys.\  {\bf 134}, 85 (1981).
\bibitem{pdv1}
P.~Di Vecchia and A.~Liccardo,
hep-th/9912161.
\bibitem{pdv2}
P.~Di Vecchia and A.~Liccardo,
hep-th/9912275.
\bibitem{mg}
M.~R.~Gaberdiel,
Class.\ Quant.\ Grav.\  {\bf 17}, 3483 (2000)
[hep-th/0005029].
\bibitem{flprs}
M.~Frau, I.~Pesando, S.~Sciuto, A.~Lerda and R.~Russo, 
Phys.\ Lett.\ B {\bf 400}, 52 (1997)
[hep-th/9702037].
\bibitem{vflprs}
P.~Di Vecchia, M.~Frau, I.~Pesando, S.~Sciuto, A.~Lerda and R.~Russo,
Nucl.\ Phys.\ B {\bf 507}, 259 (1997)
[hep-th/9707068].
\bibitem{bvflprs}
M.~Billo, P.~Di Vecchia, M.~Frau, I.~Pesando, S.~Sciuto, 
A.~Lerda and R.~Russo, 
R-R states",
Nucl.\ Phys. B {\bf 526}, 199 (1998)
[hep-th/9802088]
\bibitem{ivv1}
I.~V.~Vancea,
Phys.\ Lett.\ B {\bf 487}, 175 (2000)
[hep-th/0006228].
\bibitem{ivv2}
M.~C.~Abdalla, A.~L.~Gadelha and I.~V.~Vancea,
Phys.\ Lett.\ A {\bf 273}, 235 (2000)
[hep-th/0003209].
\bibitem{ivv3}
M.~C.~Abdalla, A.~L.~Gadelha and I.~V.~Vancea,
hep-th/0104068.
\bibitem{superD}
C.~P.~Bachas,
hep-th/9806199.
\bibitem{Schwarz:2000ew}
J.~H.~Schwarz,
hep-ex/0008017.
\bibitem{Sen:2001gy}
A.~Sen,
Nucl.\ Phys.\ Proc.\ Suppl.\  {\bf 94}, 35 (2001)
[hep-lat/0011073].
\bibitem{Schwarz:1999vu}
J.~H.~Schwarz,
hep-th/9908144.
\bibitem{Dangles}
S.~P.~de Alwis,
Phys.\ Lett.\ B {\bf 461}, 329 (1999)
[hep-th/9905080].
\bibitem{SheikhJabbari:1998cv}
M.~M.~Sheikh Jabbari,
Phys.\ Lett.\ B {\bf 420}, 279 (1998)
[hep-th/9710121].
\bibitem{Arfaei:1998hb}
H.~Arfaei and M.~M.~Sheikh Jabbari,
Nucl.\ Phys.\ B {\bf 526}, 278 (1998)
[hep-th/9709054].
\bibitem{Arfaei:1997rg}
H.~Arfaei and M.~M.~Sheikh Jabbari,
Phys.\ Lett.\ B {\bf 394}, 288 (1997)
[hep-th/9608167].
\bibitem{Townsend:1998ci}
P.~K.~Townsend,
hep-th/9901102.
\bibitem{ass1}
A.~Sen,
JHEP {\bf 9808}, 010 (1998)
[hep-th/9805019].
\bibitem{ass2}
A.~Sen,
hep-th/9904207.
\bibitem{alr}
A.~Lerda and R.~Russo,
Int.\ J.\ Mod.\ Phys.\ A {\bf 15}, 771 (2000)
[hep-th/9905006].
\bibitem{Vancea:2001zu}
I.~V.~Vancea,
JHEP {\bf 0104}, 020 (2001)
[hep-th/0011251].
\bibitem{suact}
M.~Aganagic, C.~Popescu and J.~H.~Schwarz,
Phys.\ Lett.\ B {\bf 393}, 311 (1997)
[hep-th/9610249].
\bibitem{Townsend:1999hi}
P.~K.~Townsend,
hep-th/0004039.
\bibitem{Townsend:1997wg}
P.~K.~Townsend,
hep-th/9712004.
\bibitem{geometry}H.~ B.~ Lawson,~ Jr. and M.-L.~Michelson,
{\it Spin Geometry}, (Princeton University Press, 1989) 
\bibitem{dualities}
P.~C.~West,
hep-th/9811101.
\bibitem{Duff:1999rk}
M.~J.~Duff,
hep-th/9912164.
\bibitem{tsey}
A.~A.~Tseytlin,
hep-th/9908105.
\bibitem{seibwittnc}
N.~Seiberg and E.~Witten,
JHEP {\bf 9909}, 032 (1999)
[hep-th/9908142].
\bibitem{tach}
A.~Sen,
Class.\ Quant.\ Grav.\  {\bf 17}, 1251 (2000).
\bibitem{Sen:1998sp}
A.~Sen,
{\it Prepared for 29th International Conference on High-Energy Physics 
(ICHEP 98), Vancouver, British Columbia, Canada, 23-29 Jul 1998}.
\bibitem{Sen:1999mg}
A.~Sen,
hep-th/9904207.
\bibitem{ktheory}
E.~Witten,
hep-th/0006071.
\bibitem{wittt}E.~Witten,
Int.\ J.\ Mod.\ Phys.\ A {\bf 16}, 693 (2001)
[hep-th/0007175].
\bibitem{Olsen:1999xx}
K.~Olsen and R.~J.~Szabo,
Adv.\ Theor.\ Math.\ Phys.\  {\bf 3}, 889 (1999)
[hep-th/9907140].
\bibitem{Olsen:2000dw}
K.~Olsen and R.~J.~Szabo,
Nucl.\ Phys.\ B {\bf 566}, 562 (2000)
[hep-th/9904153].
\bibitem{duliu1}
D.~Diaconescu, G.~Moore and E.~Witten,
hep-th/0005091.
\bibitem{duliu2}
D.~Diaconescu, G.~Moore and E.~Witten,
hep-th/0005090.
\bibitem{eualgk}
I.~V.~Vancea,
hep-th/9905034.
\bibitem{moore}
C.~I.~Lazaroiu,
JHEP {\bf 0106}, 064 (2001)
[hep-th/0105063].
\bibitem{Douglas:2000gi}
M.~R.~Douglas,
hep-th/0011017.
\bibitem{mald}
J.~L.~Petersen,
Int.\ J.\ Mod.\ Phys.\ A {\bf 14}, 3597 (1999)
[hep-th/9902131].
\bibitem{pvmal1}
P.~Di Vecchia,
Fortsch.\ Phys.\  {\bf 48}, 87 (2000)
[hep-th/9903007].
\bibitem{Skenderis:2000bs}
K.~Skenderis,
Lect.\ Notes Phys.\  {\bf 541}, 325 (2000)
[hep-th/9901050].
\bibitem{pvmal2}
P.~Di Vecchia,
hep-th/9908148.
\bibitem{rs}
N.~Arkani-Hamed, S.~Dimopoulos, N.~Kaloper and J.~March-Russell,
hep-ph/9903239.
\bibitem{smc}
S.~M.~Carroll,
hep-th/0011110.


\end{thebibliography}
\end{document}